\documentclass[a4paper,11pt]{article}
\pdfoutput=1 

\usepackage{jcappub} 

\usepackage[T1]{fontenc} 
\usepackage{natbib}
\usepackage{subfigure}
\usepackage[nottoc]{tocbibind}
\usepackage{booktabs}
\usepackage{multirow}
\usepackage{enumitem}
\usepackage{longtable}
\usepackage{xcolor}

\title{\boldmath Gravitational Wave mergers \\ as tracers of Large Scale Structures}

\author[a,b]{S. Libanore}
\author[c]{, M. C. Artale}
\author[d]{, D. Karagiannis}
\author[a,b]{, M. Liguori}
\author[a,b]{,\\ N. Bartolo}
\author[a,b]{, Y. Bouffanais}
\author[a,b]{, N. Giacobbo}
\author[a,b,e]{, M. Mapelli}
\author[a,b,e,f]{, S. Matarrese}

\affiliation[a]{Dipartimento di Fisica e Astronomia G. Galilei, Universit\`{a} degli Studi di Padova \\ Via Marzolo 8, 35131 Padova, Italy}
\affiliation[b]{INFN, Sezione di Padova, Via Marzolo 8, I$-$35131, Padova, Italy}
\affiliation[c]{Institut f$\ddot{\text{u}}$r Astro- und Teilchenphysik, Universit$\ddot{\text{a}}$t Innsbruck\\ Technikerstrasse 25/8, 6020 Innsbruck, Austria}
\affiliation[d]{Department of Physics and Astronomy, University of the Western Cape\\ Cape Town 7535, South Africa}
\affiliation[e]{INAF, Osservatorio Astronomico di Padova\\ Vicolo dell'Osservatorio 5, I$-$35122, Padova, Italy}
\affiliation[f]{Gran Sasso Science Institute (GSSI), Viale F. Crispi 7, I$-$67100, L'Aquila, Italy}

\emailAdd{sarah.libanore@phd.unipd.it}

\abstract{Clustering measurements of Gravitational Wave (GW) mergers in Luminosity Distance Space can be used in the future as a powerful tool for Cosmology. 
We consider tomographic measurements of the Angular Power Spectrum of mergers both in an Einstein Telescope-like detector network and in some more advanced scenarios (more sources, better distance measurements, better sky localization). We produce Fisher forecasts for both cosmological (matter and dark energy) and merger bias parameters. 
Our fiducial model for the number distribution and bias of GW events is based on results from hydrodynamical simulations.  
The cosmological parameter forecasts with Einstein Telescope are less powerful than those achievable in the near future via galaxy clustering observations with, e.g., Euclid. However, in the more advanced scenarios we see significant improvements. Moreover, we show that bias can be detected at high statistical significance.
Regardless of the specific constraining power of different experiments, many aspects make this type of analysis interesting anyway. For example, compact binary mergers detected by Einstein Telescope will extend up to very high redshifts, particularly for binary black holes. Furthermore, Luminosity Distance Space Distortions in the GW analysis have a different structure with respect to Redshift-Space Distortions in galaxy catalogues. Finally, measurements of the bias of GW mergers can provide useful insight into their physical nature and properties.}

\begin{document}
\maketitle
\flushbottom

\newpage
\section{Introduction}
\label{sec:intro}

The importance of the recent discovery of Gravitational Waves (GW) produced by Black Hole (BH) and Neutron Star (NS) mergers cannot be overemphasized  (important achievements in this new research field are described e.g., in \cite{Abbott_2016, Abbott2017, Abbott_2017_ns, Abbott_2020, LVC_1, LVC_2, LVC_3, LVC_4, LVC_5, LVC_6, LVC_7, LVC_8, LVC_9, LVC_10, LVC_11}). It has opened a new window in our understanding of the Universe, with a huge future discovery potential in many different areas of Astronomy.
If we consider the field of Cosmology, one of the most investigated applications is the use of GW events as {\it standard sirens}, to measure cosmological distances and the Hubble parameter without the calibration issues which arise in traditional approaches. This has gained even further interest in recent times, in light of the more and more debated discrepancy between measurements of the Hubble parameter, coming from high and low-redshift cosmological probes  (see e.g., \cite{LVC_10, LVC_11, Schutz1986, Abbott2017, Chen2018}). 
One caveat is that this methodology requires spectroscopic follow-ups, electromagnetic counterparts or cross-correlation of the GW signal with external galaxy surveys, in order to determine redshifts of the GW events. 

A logical question therefore arises, namely whether we can extract useful cosmological information from future GW observations, without any additional redshift information. Future GW experiments, such as Einstein Telescope (ET)\footnote{\texttt{http://www.et-gw.eu}} or DECIGO\footnote{\texttt{https://decigo.jp/index$\_$E.html}}, will detect hundreds of thousand or millions of events. Therefore, an interesting possibility is that of using GW mergers as tracers of Large Scale Structures (LSS), in essentially the same way as done with galaxies in big cosmological surveys. This does not necessarily require knowledge of redshifts, since luminosity distances -- which are directly measured -- can be used as radial coordinates. 
Using luminosity distances introduces also another layer of complementarity with galaxy surveys, since distortions of the merger distribution in Luminosity Distance Space behaves differently from distortions of the galaxy distribution in Redshift Space.

It is also interesting to point out that statistical studies of the spatial distribution of GW events allow us to characterize their clustering properties, with respect to the underlying Dark Matter (DM) distribution, i.e., their cosmological bias. From an observational point of view, studies of the spatial distribution of mergers have already been carried on in  \cite{stiskalek_2020, payne_2020}, where it was shown that GW produced by binary BH mergers are anisotropically distributed. Attempts at measuring their correlation function and power spectrum are also ongoing, see e.g. \cite{banagiri_2020}.
Modelling merger bias is important when seeking cosmological information, since in this case bias parameters need to be marginalized out in the analysis.
Beyond this aspect, bias measurements could also directly provide interesting information on the physical nature of the different mergers. Such approach is for example explored in \cite{Raccanelli_2016, Scelfo_2018, Calore_2020}. In those works, merger bias is studied via cross-correlation between galaxy and GW surveys, rather than by relying on GW experiments alone. An approach to measuring GW bias, which relies solely on source-location posteriors, has been instead proposed in \cite{vijaykumar_2020}. While we were in the final stages of this work, a new method to precisely infer redshifts of mergers and to estimate cosmological and bias parameters, without identifying their host galaxy, was also discussed in \cite{mukherjee2020accurate}; this approach extends the technique originally developed in \cite{Mukherjee_2018} for Supernovae catalogues. 
The possibility of building surveys of the spatial distribution of GW mergers -- and use them for cosmological applications -- without relying on external data, but working directly in Luminosity Distance Space, was instead originally pointed out in \cite{Namikawa_2016, Pengjie_2018}. 

In this work we go beyond these preliminary studies, by systematically exploring this approach both for a network of ET-like detectors and for more futuristic scenarios. We produce detailed Fisher forecasts for cosmological parameters (describing matter and dark energy) in all different cases, and in doing so we do not rely on simplified analytical assumptions. 
In particular, we use the results from \cite{Artale_2018,Artale_2020} to model the expected density of mergers in the survey and to characterize their fiducial bias parameters via a simulation-based Halo Occupation Distribution (HOD) approach.
The work from \cite{Artale_2018,Artale_2020} combines galaxy catalogs from hydrodynamical cosmological simulations together with the results of population synthesis models. In this way, the merger rates are computed considering galaxy and binary stellar evolution in a self-consistent way.

As mentioned above, a potentially interesting application is that of focusing on the bias parameters and trying to use them to extract information on type and properties of the mergers. We will therefore also provide specific forecasts on bias, after marginalizing over cosmological parameters (with and without priors from external cosmological surveys).

The paper is structured as follows: in Sec. \ref{sec:angsurveys} we compare (angular) merger and galaxy surveys, discussing in particular the use of luminosity distances as position indicators and the related Luminosity Distance Space Distortions; in Sec. \ref{sec:sources} we study the number distribution of events and describe our method to produce a fiducial model for merger bias; in Sec. \ref{sec:forecasts} we provide details on our Fisher matrix implementation; in Sec. \ref{sec:results} we illustrate our results. We then draw our conclusions in Sec. \ref{sec:conclusions}.

\section{Luminosity Distance Space}\label{sec:angsurveys}
This work aims at understanding how well future surveys of GW mergers will be able to constrain either Cosmology or the statistical properties of their distribution (let us note here that we focus on merger clustering in this work, but lensing studies are of course also possible and interesting, see e.g. \cite{Mukherjee_2020,Mukherjee_2020_lensing}). Only GW events caused by the merger of compact binaries are considered in our current analysis, i.e. systems formed by two Neutron Stars, two stellar Black Holes or one Black Hole and one Neutron Star. 
The approach we consider consists in studying the spatial clustering of mergers on large scales using their Angular Power Spectrum, pretty much in the same way as done for galaxy surveys (e.g. \cite{Euclid_2011}), despite the different astrophysical properties of the tracers.

The main difference between galaxy and merger surveys lies in the fact that for the former we measure redshifts $z$, whereas for the latter we have direct access only to luminosity distances $D_L$, which can be extracted by combining information on the strain of the gravitational signal and its frequency. Even if the redshift associated with the GW event could be extracted from external datasets, one of our goals in this work is to rely only on GW measurements. 

The use of $D_L$ instead of $z$ in mapping the source tomographic distribution requires the introduction of some corrections, which are described in Sec. \ref{sec:LDSD}. Once these are considered, the study of the power spectrum in Luminosity Distance Space (LDS) results to be completely analogous to the standard one in Redshift Space (RS). To keep the notation more familiar to the reader and more similar to the one used in LSS analysis, quantities in this work are generally expressed through their $z$-dependence, except when the $D_L$-dependence must be made strictly explicit. Remember however that, whenever we report cosmological observables as $z$-dependent in our notation, this implies a further $z(D_L)$ dependence, computed through 
\begin{equation}\label{eq:link_DLcomoving}
D_L = \frac{\chi(a)}{a} = (1+z)\int_0^z\frac{c}{H(z)} \ ,
\end{equation}
where $\chi(a)$ is the comoving distance, $a$ is the scale factor, $c$ is the speed of light and $H(z)$ is the Hubble parameter. 
Throughout this paper, whenever an explicit evaluation of eq. (\ref{eq:link_DLcomoving}) is required, we assume, if not differently specified, the fiducial cosmological parameters measured by {\it Planck 2018} \cite{Planck_2018} and reported in Tab. \ref{tab:fiducial_cosmology} in Appendix \ref{app:APS}. 

\subsection{Luminosity Distance Space Distortions}\label{sec:LDSD}
When studying the Universe in RS, peculiar velocities alter the observed position in the sky, generating Redshift-Space Distortions (RSD, see e.g. \cite{Lewis_2011}). Since in this work the mapping is done in LDS, we need to consider instead the analogous effect of Luminosity Distance Space Distortions (LDSD). In this Section, we do this by working in plane parallel approximation and we discuss in detail the derivation of a luminosity distance analogous of the Kaiser formula; our final result reproduces the formula originally shown in \cite{Pengjie_2018}.
Before proceeding with the discussion, let us note that future GW experiments will cover a large fraction of the sky; therefore, for future high precision analyses, we should actually also take into account wide-angle contributions to $D_L$,  due to volume, velocity and ISW-like effects. This will particularly matter for advanced experiments with very low instrumental error in the determination of distances, such as e.g., DECIGO (see \cite{Bertacca_2018}). The plane parallel approximation is however fully adequate for the accuracy requirements of the Fisher analysis we carry on here (which is also mostly focused on an ET-like survey, where instrumental errors tend to dominate over other effects in affecting measurements of $D_L$).  

The way peculiar velocities affect the observed position $D_L^{obs}$ in LDS depends both on the change in the observed position and on the relativistic light aberration. A first-order derivation (\cite{pyne_1996, pyne_2004}, see also  \cite{Lam_2006}) leads to the expression:
\begin{equation}\label{eq:obs_DL}
D_L^{obs} = \bar{D}_L(1+2\vec{v}_e\cdot\hat{n}) \ , 
\end{equation}
where $\bar{D}_L$ is the luminosity distance in the unperturbed background, $\vec{v}_e$ is the peculiar velocity of the emitting source and $\hat{n}$ is the Line of Sight (LoS) direction. 

\begin{figure}[t!]
\centering
\includegraphics[width = 0.9\textwidth]{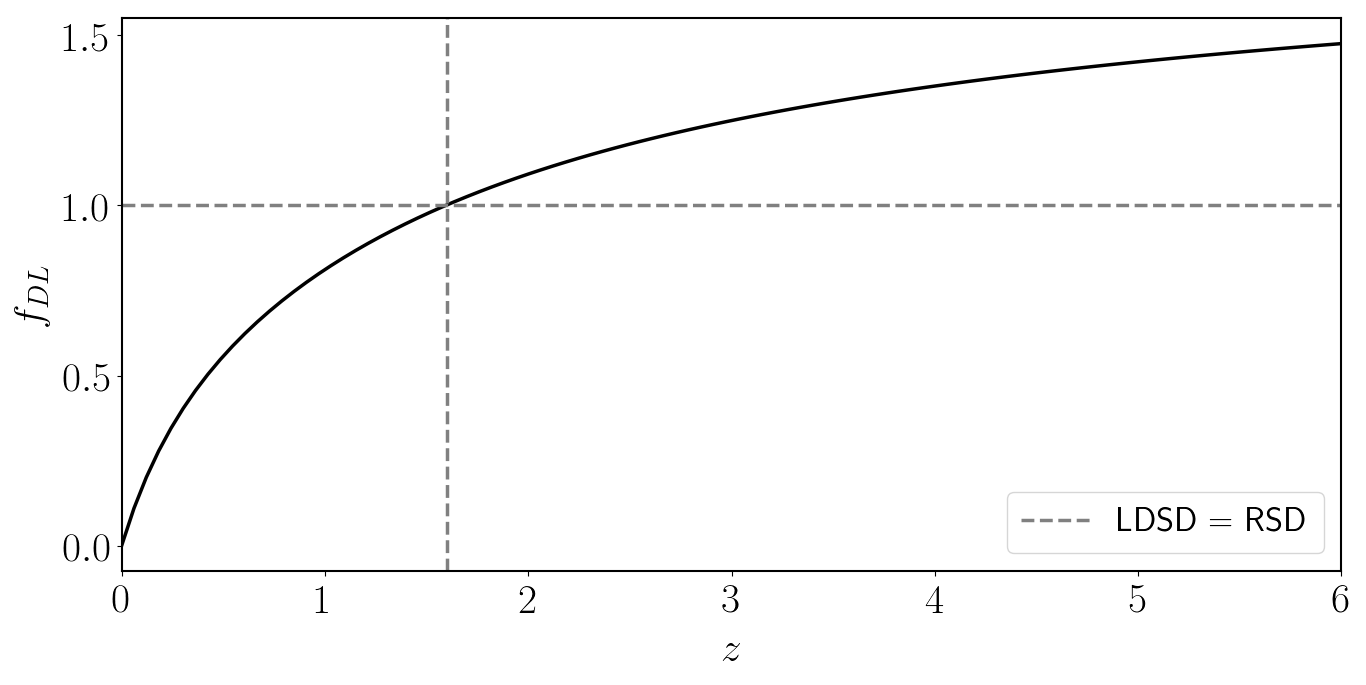}
\caption{$f_{D_L}$ factor calculated in eq. (\ref{eq:fDL}) assuming the fiducial Cosmology (see Tab. \ref{tab:fiducial_cosmology}). The dotted lines indicate the point in which LDSD are equivalent to RSD, that is $z \in [1.6, 1.7]$. Below this value, LDSD are smaller than RSD and $f_{D_L}$ varies quite fast. Differently, over it LDSD start taking over RSD while $f_{D_L}$ tends to become constant.}
\label{fig:fDL}
\end{figure}

As mentioned above, eq. (\ref{eq:obs_DL}) is used in \cite{Pengjie_2018} to describe the LDSD in a flat Universe, adopting the plane-parallel approximation, namely: 
\begin{equation}\label{eq:par_plane}
\vec{v}_e\cdot\hat{n} = \mu v_e\ .
\end{equation} 
In the previous equation, $\mu$ is the cosine of the angle between the LoS direction and the peculiar velociy of the source.
Background coordinates in real space are associated to coordinates in LDS by means of eq. (\ref{eq:link_DLcomoving}), leading to $\chi(D_L^{obs}) = aD_L^{obs} = a\bar{D}_L + a\delta D_L$. Considering eq. (\ref{eq:obs_DL}) and replacing the approximation from eq. (\ref{eq:par_plane}), we get:
\begin{equation}\label{eq:obs_chi}
\chi(D_L^{obs}) = a\bar{D}_L(1+2\mu v_e) = \chi(\bar{D}_L)(1+2\mu v_e)\ .
\end{equation}
Therefore, $\delta D_L = 2a\bar{D}_L\mu v_e$. Eq. (\ref{eq:obs_chi}) can be rewritten as:
\begin{equation}\label{eq:expand_obs_chi}
\chi(D_L^{obs}) = \chi(\bar{D}_L + \delta D_L) = \chi(\bar{D}_L) + \frac{\partial \chi(D_L^{obs})}{\partial z}\biggl(\frac{\partial D_L^{obs}}{\partial z}\biggr)^{-1}\biggl|_{\bar{D}_L}\delta D_L\ .
\end{equation}
Writing $\delta D_L$ explicitly and considering that $\delta \chi / \delta z = 1/H(z)$ in a spatially flat Universe:
\begin{equation}\label{eq:LDSD}
\begin{aligned}
\chi(D_L^{obs}) &= \chi(\bar{D}_L) + \frac{1}{H(z)}\biggl( \frac{\partial D_L^{obs}}{\partial z}\biggr)^{-1}\biggl|_{\bar{D}_L}2\mu v_e a\bar{D}_L \\
&= \chi(\bar{D}_L) + \biggl[\frac{2\bar{D}_L}{1+z}\biggl(\frac{\partial \bar{D}_L}{\partial z}\biggr)^{-1}\biggr]\frac{\vec{v}_e\cdot\hat{n}}{H(z)}\ .
\end{aligned}
\end{equation}
Eq. (\ref{eq:LDSD}) is identical in structure to the standard Kaiser formula in RS \cite{Kaiser_1987}. The only difference between the two is the pre-factor $f_{D_L}$, 
\begin{equation}\label{eq:fDL}
f_{D_L} = \frac{2\bar{D}_L}{1+z}\biggl(\frac{\partial \bar{D}_L}{\partial z}\biggr)^{-1}, 
\end{equation}
which was originally pointed out in \cite{Pengjie_2018}.\footnote{Note that lensing contributions to LDSD are neglected here, following \cite{Pengjie_2018}, where it is argued that they should be subdominant with respect to the peculiar velocity part.
As this work was being completed, reference \cite{namikawa_2020} appear on the arXiv. It is a more advanced study on LDSD, which explicitly includes lensing terms. We plan on accounting for such terms in future developments of our analysis.} This factor depends on the distance from the observer, and makes LDSD larger than RSD at $z \gtrsim 1.7$ and smaller than RSD at $z \lesssim 1.7$; due to this prefactor, LDSD are also vanishing as $z \rightarrow 0$, as Fig. \ref{fig:fDL} shows. Note that $f_{D_L}$ depends on Cosmology.

Eq. (\ref{eq:LDSD}) can be used to study LDSD in Fourier space, as done for RSD. Let us briefly review the standard procedure. The observed overdensity is computed through eq. (\ref{eq:LDSD}) and Fourier transformed
(note that, in the transform, the source redshift $\bar{z}$ is fixed, when considering the spatial distribution of the velocities; the background $\bar{D}_L$ therefore depends only on $\bar{z}$ and not on the LoS direction $\hat{n}$).
We then use the continuity equation:  
\begin{equation}\label{eq:continuity}
\dot{\delta}_k(\eta) + ikv_k(\eta) = 0\ ,  
\end{equation}
where $\eta$ is the conformal time, and use it to express the velocity as:
\begin{equation}\label{eq:vk}
v_k(\eta) = \frac{i}{k}\frac{d\delta_k(\eta)}{d\eta} = \frac{i}{k}\frac{d}{d\eta}\biggl[\frac{\delta_k(\eta) D_1(\eta)}{D_1(\eta)}\biggr]
=  \frac{i\delta_k(\eta)}{kD_1(\eta)}\frac{dD_1(\eta)}{d\eta}\ .
\end{equation}
In eq. (\ref{eq:vk}), the last equality descends from $\delta_k(\eta)/D_1(\eta) \sim cost$, $D_1(\eta)$ being the growth factor. The dimensionless growth rate is then defined as:
\begin{equation}\label{eq:f}
f = \frac{a}{D_1}\frac{dD_1}{da} = \frac{a}{D_1}\frac{1}{a^2H}\frac{dD_1}{d\eta} = \frac{1}{aHD_1}\frac{dD_1}{d\eta}\ ,
\end{equation}
where the $\eta$ dependence is omitted for clarity. Therefore, eq. (\ref{eq:vk}) is rearranged as:
\begin{equation}\label{eq:vk_f}
v_k = \frac{ifaH\delta_k}{k}\ .     
\end{equation}

\noindent Moving to LDS, the factor $f$ as reported in eq. (\ref{eq:f}) has now to be converted into $f_1 = f\cdot f_{D_L}$, with $f_{D_L}$ from eq. (\ref{eq:fDL}).

\subsection{Numerical implementation}\label{sec:LDSD_implement}
Sec. \ref{sec:LDSD} shows that LDSD, in the plane-parallel approximation, can be formally treated as done for RSD, once the factor $f_{D_L}$ from eq. (\ref{eq:fDL}) is properly inserted. Consequently, such factor enters the Angular Power Spectrum (APS) computation.

\noindent The density contrast of the sources can be written (see e.g. \cite{Lewis_2011}) as:
\begin{equation}\label{eq:transfer_real}
\delta_N = \delta_{\mathcal{N}} -\frac{1}{\mathcal{H}}\hat{n}\cdot\nabla(\vec{v}_e\cdot\hat{n}) + A(\vec{v}_e\cdot\hat{n}) + ...\ ,
\end{equation}
where the first term is the proper number density contrast at the source, the second represents the RSD/LDSD and the last one is due to the Doppler effect. Other observational effects are neglected in this expression but can be found in \cite{Lewis_2011}. 

By Fourier transforming eq. (\ref{eq:transfer_real}), the theoretical transfer function $\Delta_l(z,k)$ is obtained. The observational transfer function 
$\Delta_{N,l}^W(z,k)$ is then computed: it accounts for the redshift dependence of the source distribution $p(z)$ and for a suitable weight in each observed redshift bin provided by the Window function  $W(z_i,z)$ (see Sec. \ref{sec:APS} and e.g. \cite{Fonseca_2019} for details). 

When we compute the transfer function in LDS, each term including $v_k$ in $\Delta_l(z,k)$ inherits the factor $f_{D_L}$ from eq. (\ref{eq:fDL}). 
Therefore, such modifications are inserted in the terms describing the Space Distortions, the density evolution and the Doppler effect. Following \cite{Lewis_2011}:

\begin{equation}\label{eq:RS_to_LDS}
\begin{aligned}
\text{RSD} \sim k v_k\ j_l''(k\eta) &\to \text{LDSD} \sim f_{D_L}k v_k\ j_l''(k\chi)\ ,\\
\text{RS evolution} \sim v_k\ j_l'(k\chi) &\to \text{LDS evolution} \sim f_{D_L}v_k\ j_l'(k\chi)\ , \\
\text{RS Doppler} \sim  v_k\ j_l'(k\chi) &\to \text{LDS Doppler} \sim  f_{D_L}v_k\ j_l'(k\chi)\ .
\end{aligned}
\end{equation}
\\
In this work, the APS is computed using the public code \texttt{CAMB}\footnote{\texttt{https://github.com/cmbant/CAMB}}, introduced in \cite{Lewis_2002}. When calculating the APS in RS, the code relies on the integrated version of the expressions in eq. (\ref{eq:RS_to_LDS}), which all depend on the spherical Bessel function $j_l(k\chi)$ and not on its derivatives. 
The conversion to LDS is simple if we consider a sufficiently fine distance binning of the data, when computing the APS. In this case, without loss of accuracy, we can neglect the dependence of $f_{D_L}$ on $\chi$, inside any given bin. By doing so, \texttt{CAMB} built-in expressions are simply multiplied by $f_{D_L}$, which is computed through eq. (\ref{eq:fDL}) in the centre of the bin.

\section{Source properties}\label{sec:sources}
If we want to study the clustering of GW merger events, both their number distribution in redshift and their bias with respect to the underlying smooth DM distribution need to be modelled. To this purpose, we rely on simulations from \cite{Artale_2018,Artale_2020}. 
These combine the galaxy catalogue from the {\sc eagle} simulation \cite{Schaye2015} with the stellar population synthesis code {\sc mobse} \cite{Giacobbo2018} to get the number distribution of mergers from Double Neutron Stars (DNS), Double Black Holes (DBH) and Black Hole Neutron Star (BHNS) systems.\footnote{In this work, when talking about \emph{distributions}, binary mergers or GW events are considered interchangeably, since the former triggers the latter. The \emph{distributions} are ET selected, unless specified otherwise.} These distributions depend on the redshift $z$, the star formation rate $SFR$ and the stellar mass of the host galaxy $M_*$. Other dependencies such as the one on metallicity, are neglected for simplicity in this work, as they are found elsewhere to be subdominant in determining the rate of events \cite{Artale_2018, Artale2_2019}. Moreover, as we show in the remainder, our final merger bias predictions are consistent with, e.g., those of \cite{viel}, where a different, semi-analytical bias model is considered, in which stellar mass is neglected, while metallicity is included.  
The full distributions are finally processed to include observational effects from ET. More details about the simulations are provided in Appendix \ref{app:sim}.

\subsection{Number distribution}\label{sec:num_dist}
Simulations are run in a box having a comoving side $\ell = 25  \text{Mpc}$, which is evolved across cosmic time. Even if the box is small, for our purposes this does not generate sample variance related problems. We checked this by comparing relevant results for our analysis with similar figures obtained from a simulated box with size $\ell'= 100 \text{Mpc}$ and verifying their stability.

The simulation is divided into $22$ redshift snapshots, in which the number distributions of both galaxies and DNS/DBH/BHNS mergers are calculated. Starting from the total number of events, a filtering procedure is then applied to select only the sources which are expected to be observable with an ET-like instrument (see \cite{Baibhav_2019} for details on this procedure). Note that each $z$-snapshot actually corresponds to $[z - \delta z, z+\delta z] = T^{SIM}$; the center of each snapshot and the associated interval in units of time are reported in Tab. \ref{tab:snap} in Appendix \ref{app:sim}.

The number distribution of mergers inside the box depends in our model not only on redshift but also on the stellar mass of the host galaxy, $M_*$ and on the star formation rate, $SFR$. Both $M_*$ and $SFR$ are divided into $15$ bins, which are reported in Tab. \ref{tab:stellar_mass} in Appendix \ref{app:sim}. For fixed redshift $z$, we then sum over all the $[M_*,SFR]$ bins, in order to obtain a final merger distribution which depends only on $z$. 

In this work, we derive separate, independent forecasts for two kind of mergers, namely DBH and DNS; forecasts from BHNS would provide intermediate results between the two (less constraining than DBH, more constraining than DNS). A multitracer analysis, including all types of mergers in a single forecast, is left for a forthcoming analysis.

The distributions considered are:
\begin{equation}
N_m^{SIM}(z) = \sum_{i}\sum_{j} \bigl<N^{SIM}_m(z)|M_*^{i},SFR^j\bigr> \ ,
\end{equation} 
where $m =$ DBH, DNS.
$N_m^{SIM}(z)$ indicates the number of DBH/DNS binaries that merge inside the box of comoving volume $V^{SIM}$ in a given time interval $T^{SIM}(z)$ (see Tab. \ref{tab:snap}).
Therefore, the merger rate of these events is $N^{SIM}_m(z)/T^{SIM}(z)$. This can be transformed into a detection rate by converting time intervals from the source to the observer rest frame. 

\noindent The conversion factor is $ 
dt^{SIM}/dt^{OBS} = 1/(1+z)$.    
Therefore, we get:

\begin{equation}\label{eq:tobs_tsim}
N_m(z)\ =\ T^{OBS}\ \frac{N_m^{SIM}(z)}{T^{SIM}(z)}\ \frac{dt^{SIM}}{dt^{OBS}}\ =\ T^{OBS}\ \frac{N_m^{SIM}(z)}{T^{SIM}(z)}\ \frac{1}{1+z}\ ,
\end{equation} 

\noindent where $T^{OBS}$ is the survey duration expressed in years. Here, a $3 \text{yr}$ observation run is assumed.

The final step is to convert the merger number distribution into the number density of observed mergers per unit redshift and solid angle: $d^2N_m/dzd\Omega$. The solid angle $\Delta \Omega_{box}$ under which we see the surface delimiting the simulation box, at a given redshift $z$, is:
\begin{equation}
\Delta \Omega_{box} = \biggl(\frac{D_L(z)}{\ell\ (1+z)^2}\biggr)^2,
\end{equation}
where $\ell$ is the length of the simulation box side, specified earlier.
This leads to the following formula for the merger number densities:
\begin{equation}\label{eq:dNdzdOmega}
\frac{d^2N_m}{dzd\Omega} = N_m(z)\frac{c}{\ell\ H(z)}\biggl(\frac{D_L(z)}{\ell\ (1+z)^2}\biggr)^{-2}.
\end{equation}
The values of $d^2N_m/dzd\Omega$ are computed in the $22$ snapshots of the simulation. An interpolation is then performed to the skewed Gaussian:
\begin{equation}\label{eq:dNdzdOmega_skew}
\frac{d^2N_m}{dzd\Omega} = 2\biggl[A\exp\biggl(-\frac{(z-\bar{z})^2}{2\ \sigma^2}\biggr)\biggr]\biggl[\frac{1}{2}\biggl(1+ \text{erf}\biggl(\frac{\alpha(z-\bar{z})}{\sigma^2\sqrt{2}}\biggr)\biggr)\biggl], 
\end{equation}
finding $A = 10^{3.22}, \bar{z}=0.37, \sigma^2  =1.42, \alpha = 5.48$ if $m =$ DBH, while $A = 10^{3.07}, \bar{z}=0.19, \sigma^2 = 0.15, \alpha = 0.8$ if $m =$ DNS.
Our distributions are in agreement with results from \cite{viel}, keeping into account that our observational ET selection function produces a stronger decrease, in our case, in the number of DNS compared to DBH.

\begin{figure}[t!]
\centering
\includegraphics[width = 0.9\textwidth]{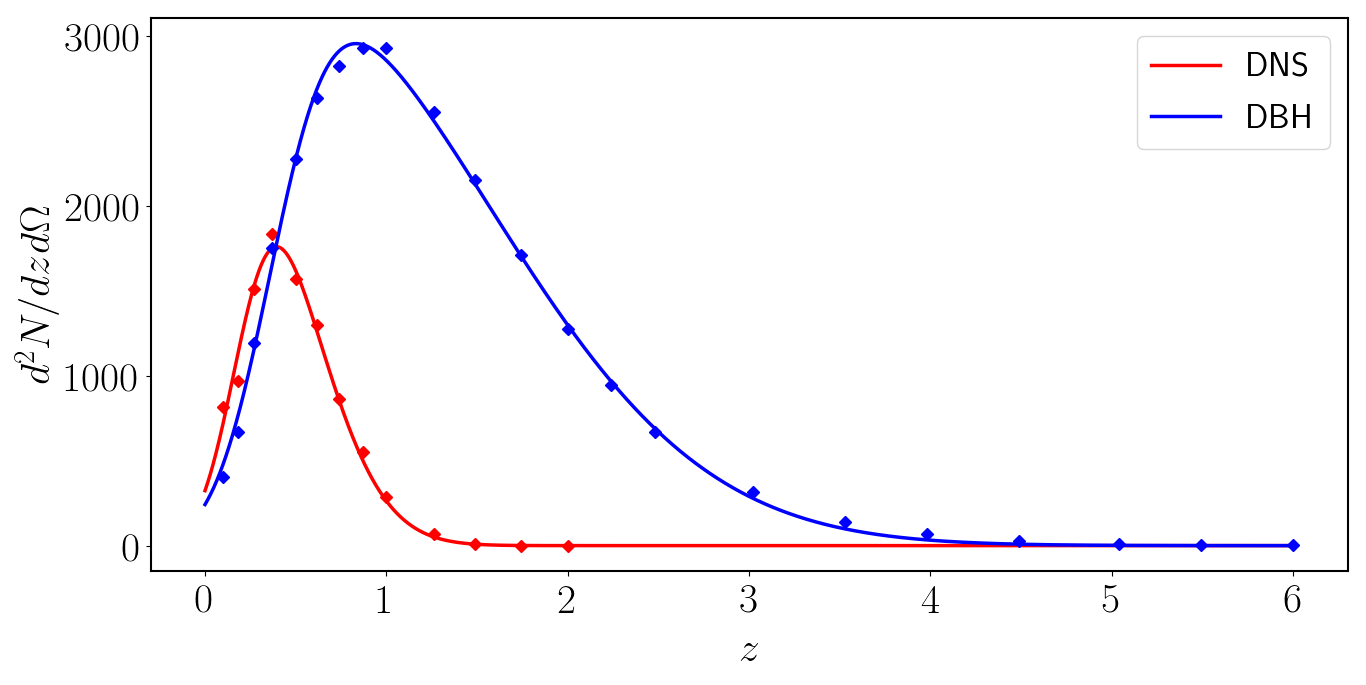}
\caption{Observed merger number distributions $d^2N_{DNS}/dzd\Omega$ (red) and $d^2N_{DBH}/dzd\Omega$ (blue) (see Sec. \ref{sec:num_dist}), for an ET-like mock survey, assuming a $3\text{yr}$ integration period. The total number of sources over the full sky in the entire redshift interval considered ($0 \leq z \leq 2$ for DNS; $0 \leq z \leq 6$ for DBH) is, for DNS, $\sim 10^{4.14}$, while for DBH it is $\sim 10^{4.79}$.}
\label{fig:dndzdomega}
\end{figure}

\subsection{Bias computation}\label{sec:bias}

The method used to get the merger bias is based on the HOD approach. This is commonly used to compute the bias of a particular kind of galaxies depending on the probability that a certain number of them form inside a DM halo having mass $M_h$ (see e.g. \cite{Cooray_2002}). Since mergers take place inside galaxies, an extra layer is added in the computation to link the merger distribution properties to the galaxy distribution (and consequently to DM, via galaxy bias). Specifically, merger bias is computed as:

\begin{equation}\label{eq:biasmerger}
b_m(z) = \int_{M_*^{min}}^{M_*^{max}} dM_*\ \int_{SFR^{min}}^{SFR^{max}} dSFR\ n_g(z,M_*,SFR)\ b_g(z,M_*,SFR)\frac{\bigl<N_m(z)|M_*,SFR\bigr>}{n_m(z)} \ .
\end{equation}

The merger HOD $\bigl<N_m(z)|M_*,SFR\bigr>$ is extracted from the simulations described in Sec. \ref{sec:num_dist}; it is used to compute the merger number density as:

\begin{equation}\label{eq:mer_den}
n_m(z) = \int_{M_*^{min}}^{M_*^{max}} dM_*\ \int_{SFR^{min}}^{SFR^{max}} dSFR\ n_g(z,M_*,SFR)\ \bigl<N_m(z)|M_*,SFR\bigr>\ .     
\end{equation}

\noindent As for the other quantities in eq. (\ref{eq:biasmerger}):
\begin{equation}
n_g(z,M_*,SFR)= \int_{M_h^{min}}^{+\infty} dM_h\ n_h(z,M_h)\ \bigl<N_g(M_*,SFR)|M_h\bigr>
\end{equation} is the mean number density of galaxies having stellar mass $M_*$ and star formation rate $SFR$, while $b_g(z,M_*,SFR)$ is their bias, again computed through a standard HOD procedure as:

\begin{equation}\label{eq:biasgal_m}
b_g(z,M_*,SFR) = \int_{M_h^{min,(*,SFR)}}^{+\infty} dM_h \ n_h(z,M_h) \ b_h(z,M_h)\frac{\bigl<N_g(M_*,SFR)|M_h\bigr>}{n_g(z,M_*,SFR)}\ .
\end{equation}

\noindent In eq. (\ref{eq:biasgal_m}), $\bigl<N_g(M_*,SFR)|M_h\bigr>$ is the galaxy HOD, i.e. the number of galaxies of stellar mass $M_*$ and star formation rate $SFR$ formed inside a halo with given mass $M_h$. The minimum mass $M_h^{min,(*,SFR)}$ required from a halo to form galaxies of such stellar mass and star formation rate, is a free parameter; the procedure we adopt to find its value is described in Sec. \ref{sec:getgalbias}. Instead, $n_h(z, M_h) = dn_h/dM_h$ is the halo mass function and $b_h(M_h, z)$ is the halo bias. In this work, we adopt the Tinker et al. prescription \cite{Tinker_2008} for the halo mass function, and compute the bias as
\begin{equation}
b_h(z,M_h) = 1 + \dfrac{1}{\sqrt{a}\delta_c}\biggl[\sqrt{a}\ a\nu^2+\sqrt{a}b(a\nu^2)^{1-c}-\dfrac{(a\nu^2)^c}{(a\nu^2)^c+b(1-c)(1-c/2)}\biggr]\ ,
\end{equation}

\noindent where $\nu = \delta_c/\sigma(z,M_h)$ is computed using the critical density for spherical collapse $\delta_c$ and the mass variance $\sigma(z,M_h)$.\footnote{We acknowledge use of the \texttt{python} library \texttt{hmf} \cite{Murray_2013} to compute the halo mass function and bias related quantities, such as the mass variance $\sigma(z,M_h)$.} The other parameters are set as $a = 0.707,\ b = 0.5,\ c= 0.6$, $M_h^{min} = 10^8\ h^{-1}M_{\odot}$,  $M_h^{max} = 10^{19}\ h^{-1}M_{\odot} $.

We note here that we have chosen an HOD-based approach to compute biases for two main reasons. On one side, it is simple but at the same time sufficiently accurate for a Fisher matrix analysis, such as the one carried on in this work. On the other side, it allows for a semi-analytical description of the bias of mergers, which can be useful for general purposes. 

A thoroughly complete description of both merger distribution and bias would of course be characterized by a much larger degree of complexity than the one displayed by our HOD. For example, it would include a specific mass-dependence for each single compact binary (see e.g. \cite{de_Mink_2016}), plus the dependence on merger formation history, channel and surrounding environment (see e.g. \cite{mandel, Schneider_2017}). Such a detailed description is not necessary for the level of accuracy required in a Fisher matrix analysis, such as the one carried out in our work. 
It would moreover be very hard to include all these factors in a simple semi-analytical model with tunable parameters, like the one developed here\footnote{Of course such factors are explicitly present in the starting simulation, but they are integrated out in the present analysis, where we summarize the merger density distributions as a function of general merger type.}. The adoption of our semi-analytical framework is however still very useful at this stage, in order to build intuition and more easily assess the impact of different variables on the final bias curve.   
Of course, in the long run the accuracy of the model will have to be further refined. To this purpose, rather than a simple Fisher matrix approach, a full Monte Carlo analysis, directly based on our mock dataset, will be necessary. This is work in progress and will be the object of a future publication.

\subsubsection{Galaxy HOD and bias}\label{sec:getgalbias}

In this Section, we provide more technical details on the procedure adopted to compute the galaxy bias, as a function of $M_*$, $SFR$ and $z$.
Firstly, the Stellar Mass Function (SMF) $\Phi(z,M_*,SFR)= d^3N/dVdM_{*}dSFR$ is defined as the number of galaxies per unit comoving volume, unit stellar mass and unit star formation rate by interpolating the data extracted from the {\sc eagle} simulation (e.g. see \cite{Artale_2018}) in the $22$ redshift snapshots reported in Tab. \ref{tab:snap}. 

Using the SMF, the galaxy number density is computed in each redshift snapshot per each stellar mass bin and star formation rate bin (see Tab. \ref{tab:stellar_mass}). This is done through:

\begin{equation}
n_g(z) = h^3\int_{M_*^{min}}^{M_*^{max}}dM_*\ \int_{SFR^{min}}^{SFR^{max}}dSFR\ \Phi(z,M_*,SFR)\ .
\end{equation}

The SMF is then compared with the HOD $\bigl<N_g(M_*,SFR)|M_h\bigr>$ to set the value of $M_h^{min,(*,SFR)}$. In this work, the {\sc eagle} HOD defined in \cite{Artale_2018} is used, that is:
\begin{align}
\bigl<N_g|M_{h}\bigr> &=
\bigl<N_g^{central}\bigl>+\bigl<N_g^{satellites}\bigl> \\ 
&= \frac{\biggl[1+\text{erf}\biggl(\dfrac{\log(M_h)-\log(M_h^{min})}{\sigma_{\log (M_h)}}\biggr)\biggr]}{2}+
\begin{cases} \biggl[\dfrac{M_h-M_h^{cut}}{M_{h,1}}\biggr]^{\alpha} &\text{if} \quad \dfrac{M_h-M_h^{cut}}{M_{h,1}} \geqslant 0 \\ \quad\quad\quad 0 &\text{otherwise}\end{cases} \nonumber .
\end{align}

\noindent The parameters $\sigma_{\log(M_h)} = 0.318,\ M_h^{cut} = 10^{11.90},\ \alpha = 1.17$ are fixed, while $M_{h,1}$ is computed as $M_{h,1} = 14.25\cdot 10^{13.32}-M_h^{cut}$, as \cite{Artale_2018} indicates.

Following \cite{Karagiannis_2018}, the value of $M_h^{min} = M_h^{min,(*,SFR)}$ is fixed in each stellar mass bin and each star formation rate bin, through the minimization of:

\begin{equation}\label{eq:HOD}
\begin{aligned}
\Delta n_g =& \ h^3\int_{M_*^{min}}^{M_*^{max}}dM_*\ \int_{SFR^{min}}^{SFR^{max}}dSFR\ \Phi(z,M_*,SFR) \ +\\
& - \int_{M_h^{min,(*,SFR)}}^{M_h^{max}}\bigl<N_g|M_{h}\bigr>\ n_h(z,M_h)\, dM_h \ . 
\end{aligned}
\end{equation}

\begin{figure}[t!]
\centering
\includegraphics[width = 0.9\textwidth]{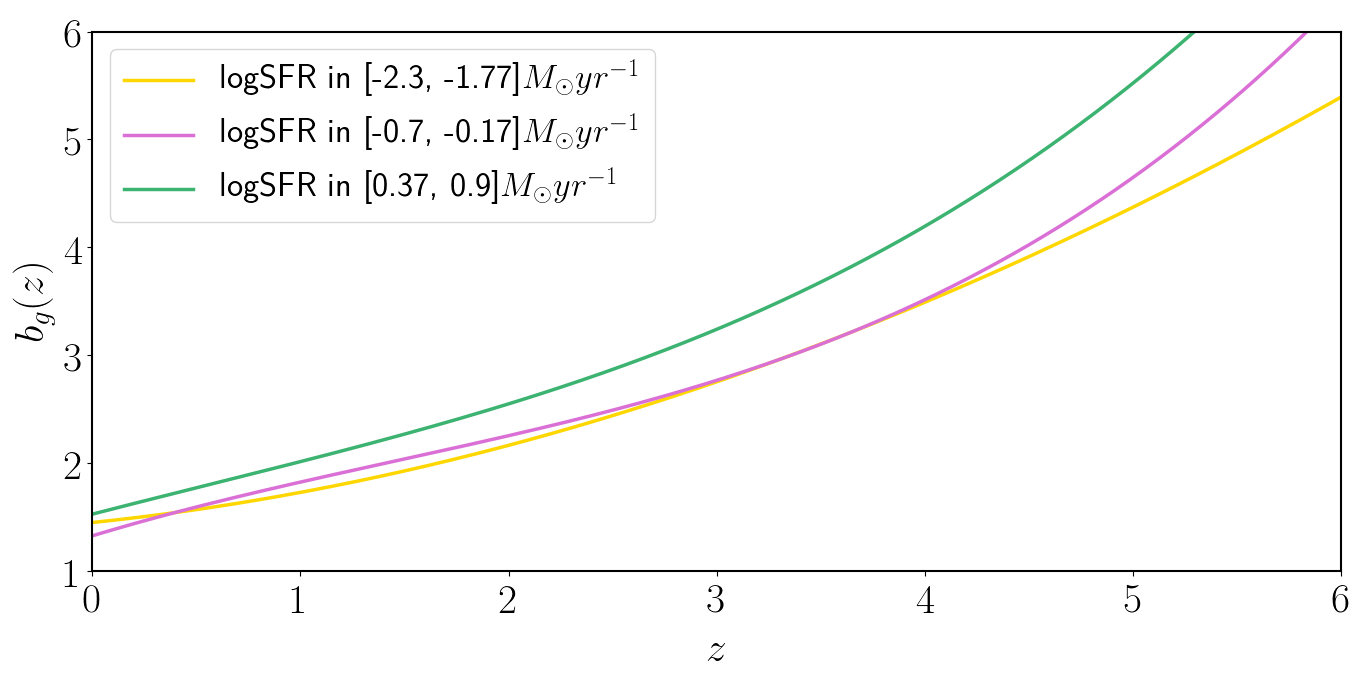}
\caption{Galaxy bias depending on $z$ in 3 different $SFR$ bin. The $M_*$ dependence has been integrated over. The yellow line shows the bias for galaxies having $SFR \in [10^{-2.3},10^{-1.77}]M_{\odot}yr^{-1}$; the purple line refers to $SFR \in [10^{-0.7},10^{-0.17}]M_{\odot}yr^{-1}$; the green line instead refers to $SFR \in [10^{0.37},10^{0.9}]M_{\odot}yr^{-1}$. All the bias models are computed by interpolating $b_g(z) = a_0+a_1z+a_2z^2+a_3z^3$ as described in eq. (\ref{eq:biasgal}).}
\label{bias}
\end{figure}

At this point, both $n_g(z,M_*,SFR)$, described in the previous Section, and the value of $b_g(z,M_*,SFR)$ can be calculated in each stellar mass bin and each star formation rate bin. The latter is computed as eq. (\ref{eq:biasgal_m}) suggests.
Fig. \ref{bias} shows some example of galaxy bias computed in different $SFR$ bins, after that the $M_*$ integration has been performed. Each curve is described by the polynomial interpolation:
\begin{equation}\label{eq:biasgal}
b_g(z)=a_0+a_1z+a_2z^2+a_3z^3 \ ,
\end{equation}
where $a_0,\ a_1,\ a_2,\ a_3$ depend on the $SFR$ bin considered. In the cases showed in the plot, they are find to be
\begin{equation*}
\begin{aligned}
& a_0 = 1.45,\ a_1 = 0.2,\ a_2 = 0.08,\ a_3 = 0.0 \ \text{if } SFR \in [10^{-2.3},10^{-1.77}] M_{\odot}yr^{-1}\ ,\\
& a_0 = 1.32,\ a_1 = 0.58,\ a_2 = -0.11,\ a_3 = 0.03 \ \text{if } SFR \in [10^{-0.7},10^{-0.17}]M_{\odot}yr^{-1} \ , \\
& a_0 = 1.52,\ a_1 =	0.5,\ a_2 = -0.03,\ a_3 = 0.02 \ \text{if } SFR \in [10^{0.37},10^{0.9}]M_{\odot}yr^{-1} \ .\\
\end{aligned}    
\end{equation*}

\subsubsection{Merger bias}\label{sec:merbias}

\begin{figure}[t!]
\centering
\includegraphics[width = 0.9\textwidth]{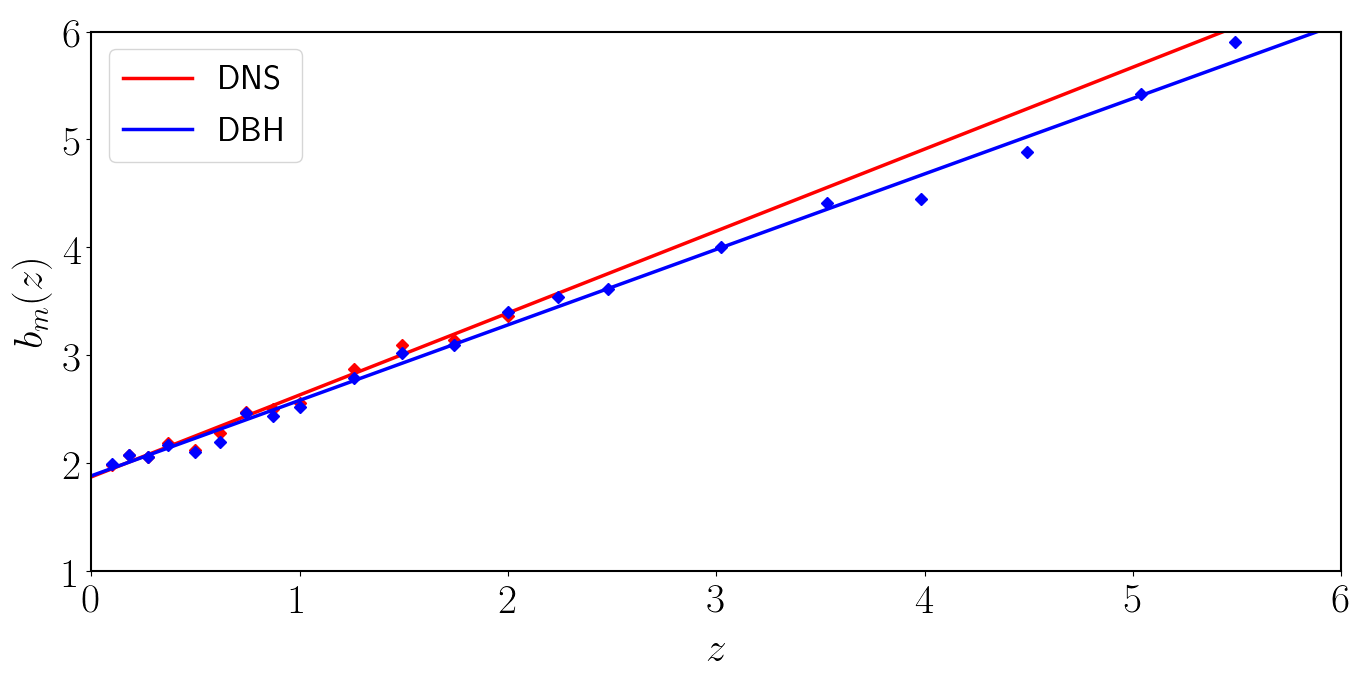}
\caption{Bias of DNS (red) and DBH (blue) distributions selected by ET. The dots indicate the values obtained in each simulation snapshot, while the line shows the linear interpolation obtained through eq. (\ref{eq:bias_m}). DNS dots stops at $z\sim 2$ in agreement with the distribution shown in Fig. \ref{fig:dndzdomega}.}
\label{fig:merbias}
\end{figure}

All ingredients are now available to compute the bias of mergers, following the prescription of eq. (\ref{eq:biasmerger}),
where the values of $n_g(z,M_*,SFR)$ and $b_g(z,M_*,SFR)$ are obtained as described in Sec. \ref{sec:getgalbias}, while $\bigl<N_m(z)|M_*,SFR\bigr>$ is derived from simulations as outlined in Sec. \ref{sec:num_dist}; finally, the merger number distribution $n_m(z)$ is calculated according to eq. (\ref{eq:mer_den}).
As Fig. \ref{fig:merbias} shows, for both DBH and DNS the bias of the mergers is well described by a linear dependence on redshift:
\begin{equation}\label{eq:bias_m}
b_m(z) = Az + B \ ,
\end{equation}
finding $A = 0.7,\ B = 1.88$ for DBH, while $A = 0.76, B = 1.87$ for DNS. 

A linear merger bias is actually often assumed in the few studies on the subject, which are currently in the literature (see e.g. \cite{Calore_2020}). In this work, we have not made any initial assumption, but we have instead explicitly worked out and justified such linear behaviour through the standard HOD approach, starting from astrophysically motivated simulations of the merger distribution. The linear trend of our bias curve is in full agreement with, e.g., the results from \cite{viel} at redshift $z < 4$. At higher redshift ($z > 4$), the results in \cite{viel} display a flattening of the bias curve, which we do not actually see. 
This is likely due to the fact that the semi-analytical model of \cite{viel} includes only mergers from star-forming galaxies, representing only a subset of the total amount of galaxies that we instead consider in the {\sc eagle} simulation. While it is in general interesting to further investigate the bias contributions from mergers at different high redshift galaxies, this does not bear any impact on the present work, since the low abundance of mergers at $z > 4$ makes their contribution to the final constraints negligible.

\section{Forecasts}\label{sec:forecasts}

Future surveys will measure the distribution of the mergers depending on both their luminosity distance and their sky position. As mentioned earlier, through these data we will be able to constrain both cosmological and 
merger bias parameters, without relying on any external measurement.
In this work, we forecast the constraining power of a 3rd-generation network of ET-like detectors and we consider also more advanced scenarios, using the Fisher formalism applied to the APS of the mergers.

\subsection{Angular Power Spectrum}\label{sec:APS}
The APS is defined as the harmonic transform of the correlation function between observed sources and it is linked to the primordial $3$D power spectrum $P^{pr}(k)$ through the standard formula:
\begin{equation}\label{eq:APS}
C_l(z_i,z_j) = 4\pi\int d\ln k \ \Delta_{N,l}^W(z_i,k)\Delta_{N,l}^W(z_j,k)\ P^{pr}(k) \ ,
\end{equation}
where $(z_i,z_j)$ are the central points of the redshift bins in which the APS is calculated, while $\Delta_{N,l}^W(z_{i,j},k)$ are the observed transfer functions in such bins, already mentioned in Sec. \ref{sec:LDSD_implement}. These are defined as:
\begin{equation}\label{eq:TF}
\Delta_{N,l}^W(z_{i},k) = \int_{z_{i}^{min}}^{z_{i}^{max}}dz\ p(z)\ W(z_{i},z)\ \Delta_l(z,k) \ ,
\end{equation}
where $W(z_{i},z)$ is the Window function considered in the redshift bin centered in $z_{i}$, and $p(z)$ is the background source distribution per redshift and solid angle. This is proportional to $d^2N_m/dzd\Omega$ but it is normalized in the bin through $\int dz\ p(z) W(z_i,z) = 1$. The full expression of the theoretical transfer function $\Delta_l(z,k)$ can be found e.g. in \cite{Lewis_2011}.

The computation of the DBH/DNS distribution $d^2N_m/dzd\Omega$ and of the bias $b_m(z)$ is discussed in Sec. \ref{sec:num_dist} and Sec. \ref{sec:bias}: eq. (\ref{eq:dNdzdOmega_skew}) and eq. (\ref{eq:bias_m}) are implemented in \texttt{CAMB} -- together with the LDSD modifications described in Sec. \ref{sec:LDSD_implement} and computed using the factor $f_{D_L}$ defined in eq. (\ref{eq:fDL}) -- to numerically compute the required APS. 

\subsubsection{Bin definition}\label{sec:bins}
To study the APS, a binning in $D_L$ is defined and converted into $z(D_L)$, after choosing fiducial values for the cosmological parameters (see Tab. \ref{tab:fiducial_cosmology}).

The amplitude of the $D_L$ bins is chosen to reproduce the predicted ET uncertainty in measuring luminosity distances. In agreement with \cite{Zhao_2018}, we assume it to be
\begin{equation}
\frac{\Delta D_L}{D_L} = 10\% \ \text{for DBH,}\quad \frac{\Delta D_L}{D_L} = 30\% \ \text{for DNS}.  
\end{equation}
A more refined definition of the error -- which should be distance dependent and linked to the sky position and inclination of each merger -- goes beyond the accuracy level required for a Fisher matrix forecast. It will be included in our future Monte Carlo analysis. In the meantime, to compensate for such lack of detail, we stick to rather conservative assignments for our $D_L$ errors. For example, we see that our $10 \%$ relative error in $D_L$ for DBH is larger than the one forecasted by \cite{Bertacca_2018}, in the entire redshift range we take into account. 
We verify that the factor $f_{D_L}$ introduced in Sec. \ref{sec:LDSD_implement} has little variation inside each one of these bins: $\Delta f_{D_L}|_{D_{L,i}^{min}}^{D_{L,i}^{max}} \leq 0.1f_{D_L}(D_{L,i}^{min})$. The approximation $f_{D_L} \simeq cost$ in a given bin is therefore completely reasonable. 

We also analyze more optimistic and more futuristic configurations, beyond the accuracy allowed by ET. In this case we assume our errors as
\begin{equation}
\frac{\Delta D_L}{D_L} = \begin{cases}
10\% \quad \text{if} \ z < 2\\
3\% \quad \text{if} \ z \geq 2\\
\end{cases}    \text{for both DBH and DNS}.
\end{equation}
While being still conservative at low distances (this choice is dictated by the fact that further reducing the $D_L$ bin size at such redshifts just introduces numerical instabilities, without improving the overall result), this configuration significantly increases the accuracy of the $D_L$ measurement at high distances. This can be compared to a DECIGO-like survey (see e.g. \cite{Bertacca_2018}), in which the nearest events are binned together. However, note that DECIGO not only will obtain more accurate $D_L$ measurements, but also observe more sources than ET. In order to further study the effect of increasing the number of observed events, we consider a rescaling of the ET source distributions, which have been derived in eq. (\ref{eq:dNdzdOmega_skew}); we refer the reader to the discussion in Sec. \ref{sec:cosmology_res} for more details.

In both the ET-like and the futuristic cases, we set a lower distance bound at  $D_L^{min} \simeq 476 \, \text{Mpc}$, which corresponds to $z^{min} = 0.1$ in the fiducial Cosmology (see Tab. \ref{tab:fiducial_cosmology}). This lower limit is chosen to stabilize the number of bins at low redshift, without any loss of cosmological information.
The highest redshift bin is chosen by considering the merger distribution, shown in Fig. \ref{fig:dndzdomega}. For DBH, since we have  $d^2N_{DBH}/dzd\Omega \simeq 0$ at $z\simeq 5$, we consider $D_L^{max} \simeq 47749 \, \text{Mpc}$, corresponding to $z^{max} \simeq 5$ in the fiducial Cosmology (Tab. \ref{tab:fiducial_cosmology}); for DNS instead, $D_L^{max} \simeq 15941 \, \text{Mpc}$, corresponding to $z^{max} \simeq 2$ in the fiducial Cosmology (Tab. \ref{tab:fiducial_cosmology}). The $D_L$ bins obtained are reported in Appendix \ref{app:APS} in Tab. \ref{tab:DLbin} and \ref{tab:DLbin_DECIGO} respectively, for the ET-like and DECIGO-like surveys.

\begin{figure}[t!]
\centering
\hspace{0.2cm}\subfigure{\includegraphics[width = 0.92\textwidth]{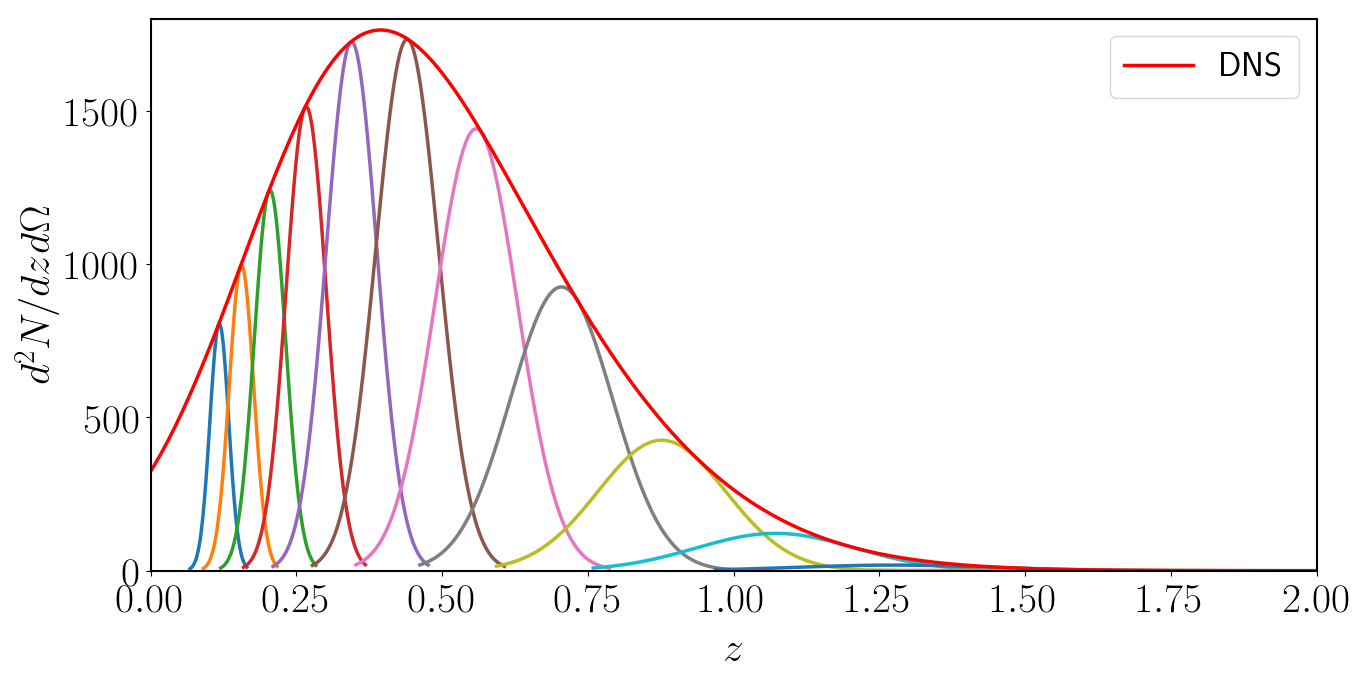}}
\subfigure{\includegraphics[width = 0.9\textwidth]{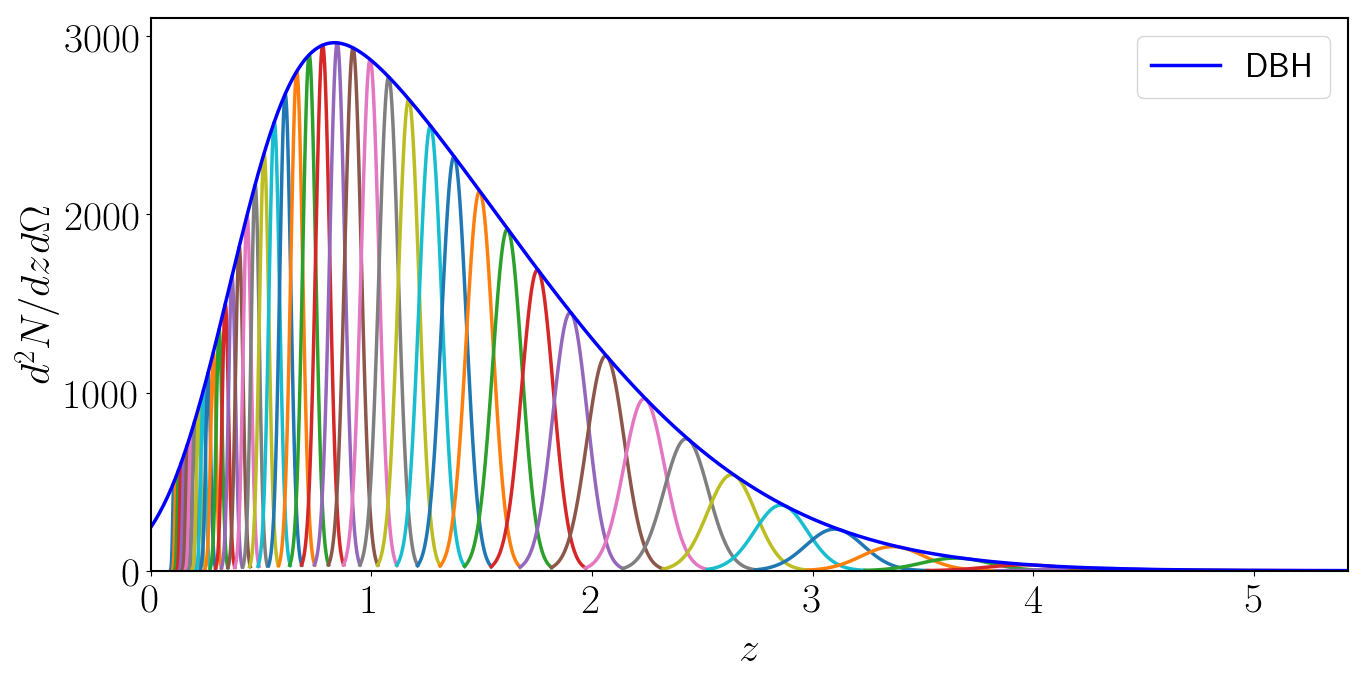}}
\caption{DNS (upper plot) and DBH (lower plot) distributions, compared with the Window functions computed in the different $z$ bins of the ET-like survey. These are computed converting the $D_L$ bins described in Sec. \ref{sec:APS} and reported in Tab. \ref{tab:DLbin} through the fiducial Cosmology (see Tab. \ref{tab:fiducial_cosmology}).}
\label{fig:DLbins}
\end{figure}

To compute the APS, a Gaussian Window function is used in each bin. This is centered in $z_i = (z_{i}^{min}+z_{i}^{max})/2$, with variance $\sigma = (z_{i}^{max}-z_{i}^{min})/2$. Fig. \ref{fig:DLbins} compares the DBH and DNS distributions from eq. (\ref{eq:dNdzdOmega_skew}), with the Gaussian Window functions in the bins in Tab. \ref{tab:DLbin}.

\subsection{Fisher matrix formalism}\label{sec:fisher}
The Fisher matrix for the APS is defined as:
\begin{equation}\label{eq:fisher}
F_{\alpha\beta} = \sum_{l_{min}}^{l_{max}}\frac{2l+1}{2}f_{sky}\sum_{D_{L,i},D_{L,j}}[(\partial_{\alpha}\mathsf{C}^{ij}_l)\ \Gamma_{l,ij}^{-1}(\partial_{\beta}\mathsf{C}^{ij}_l)\ \Gamma_{l,ij}^{-1}] \ ,
\end{equation}
where $\mathsf{C}_l$ is the APS matrix, in which $\mathsf{C}_l^{ij}=C_l(z_i,z_j)$ from eq. (\ref{eq:APS}).
The derivatives are computed with respect to the parameters of interest
\begin{equation}
\Theta = [H_0,\Omega_ch^2,w_0,w_a,b_m^0,...b_m^n]\ ,
\end{equation}
where $b_m^0,...b_m^n$ are the bias parameters for the associated merger kind $m$, defined inside each of the $D_L$ bins. The fiducial values of the cosmological parameters are reported in Tab. \ref{tab:fiducial_cosmology}, while for each bias parameter the fiducial is found through eq. (\ref{eq:bias_m}), in the central point $z_i$ of the associated bin.
We approximate the derivatives by finite differences, through the $3$-point method, with the choice $\Theta_{\alpha} = \Theta^{fid}_{\alpha} \pm 10^{-4}\Theta^{fid}_{\alpha}$, where $\Theta^{fid}_{\alpha}$ is the fiducial value of the parameter $\Theta_{\alpha}$. The APS $C_l(z_i,z_j)$ is computed by \texttt{CAMB}, using $z$ as independent variable; perturbations to account for light cone effects are already included there (for details, see \cite{Lewis_2011}). Since our bins are initially defined in LDS, there is a dependence on Cosmology in the conversion from $D_L$ to $z$, which is implicitly accounted for in the numerical derivatives. 

\noindent The term $\Gamma_{l,ij}$ in eq. (\ref{eq:fisher}) is defined as $\Gamma_{l,ij} \equiv \mathsf{C}_l^{ij}+N_l^{ij}$, $N_l^{ij}$ being the noise contribution in each bin. The amplitude of the bins defines the observational uncertainty in the determination of $D_L$, while the contribution to $N_l^{ij}$ is due to shot noise. Therefore, for each kind of merger, 
\begin{equation}\label{eq:shotnoise}
\begin{aligned}
N_l^{ij} = \delta^K_{ij} \bar{N}_{i,j}^{-1} = \delta^K_{ij}\biggl[\int_{z_{i,j}^{min}}^{z_{i,j}^{max}}\frac{d^2N_m}{dzd\Omega}\bar{W}(z_{i,j},z)\, dz \ \biggr]^{-1} ,
\end{aligned}
\end{equation}
where $\bar{W}(z_{i,j},z) = W(z_{i,j},z)/(\sigma\sqrt{2\pi})$ and $W(z_{i,j},z)$ is the Gaussian Window function. 
In eq. (\ref{eq:fisher}), $f_{sky}$ is the observed fraction of the sky -- assumed in this work to be $1$ -- while $l_{min}$ and $l_{max}$ define respectively the largest and smallest scale in the analysis. We choose $l_{min}= 2\pi/\theta = 2$ (where $\theta=\pi$ is the largest observed angular scale) and:
\begin{equation}\label{eq:lmax}
l_{max}(z_i,z_j) = k_{nl}^0(1+z_{i,j})^{2/(2+n_s)} \chi(z_{i,j})\ , 
\end{equation}

\noindent where $\chi(z_{i,j})$ is the comoving distance computed in the central point of the bin and $n_s$ is the primordial spectral index. The quantity $k_{nl}^0$ is the scale at which non-linear effects are considered too large to be properly accounted for in our approach, at $z=0$. We consider two prescriptions for this value. In the more optimistic one, we rely on the accuracy of the halofit model, which is used in \texttt{CAMB} to compute the non-linear power spectrum; therefore we include scales up to $k_{nl}^0 = 0.4\ h\text{Mpc}^{-1}$ (see \cite{Takahashi_2012}). In the more conservative one, we stick to linear scales and choose $k_{nl}^0 = 0.1\ h\text{Mpc}^{-1}$.

In the analysis, the effect of imperfect sky localization of GW events has also to be kept into account. The sky localization error, $\Delta \Omega$, smooths out fluctuations below a given scale $l_{eff}$, defined as $l_{eff}(z_i,z_j) = \chi(z_{i,j}) k_{eff}$, where:
\begin{equation}
k_{eff} = \sqrt{8\ln{2}}/(\chi(z_{i,j})\Delta\Omega^{1/2})\ .
\end{equation}
We assume a Gaussian distribution of the localization error and consider three different possibilities: 1) $\Delta\Omega = 3\ \text{deg}^2$ for DBH and $\Delta\Omega = 10\ \text{deg}^2$ for DNS (in conservative agreement with \cite{Zhao_2018}); 2) $\Delta \Omega = \ 0.5\ \text{deg}^2$ for both merger kinds and 3) a high precision localization scenario, in which $\Delta \Omega = {\rm few \, arcmin^2}$ (so that we have $l_{eff} < l_{max}$ at all redshifts).
In this work, for our baseline "ET-like" configuration we choose the first configuration. This level of localization precision, or better, is achievable with a network of three third  generation detectors, such as ET or Cosmic Explorer. Therefore, whenever we consider the "ET-like" baseline case, in our forecast, we refer to an (${\rm ET} \times 3$) network, unless otherwise specified.\footnote{An $\sim 10 \ \text{deg}^2$ localization error could also be achieved with networks of five second generation detectors; however, in our work, we consider source distributions extracted from ET mock datasets; this is why we describe our baseline scenario as a three, ET-like, detector network.
} 
Finally -- while for cosmological parameters we take $\ 10\ \text{deg}^2$ as the largest error in our analysis -- for bias parameters we also consider the case $\Delta\Omega = 100\ \text{deg}^2$, since, as we will show, this is already sufficient to achieve a bias detection.

\section{Results}\label{sec:results}
This Section reports the results of our Fisher analysis. The details of the GW survey considered are shown in Appendix \ref{app:APS} in Tab. \ref{tab:survey_details}.
Forecasts are derived for the different scenarios described in Sec. \ref{sec:bins}. For each of them, three different configurations are assumed; in each run the parameter $H_0$ is marginalized out, assuming {\it Planck 2018} results as a prior \cite{Planck_2018}. This marginalization is always carried out because data in luminosity distance space display poor constraining power on $H_0$, which appears as an overall normalization parameter after differentiating the $z(D_L)$ $H_0$-dependence, leading to degeneracies (in similar fashion as it happens e.g. in Supernovae Ia analyses).
In "run A", we fix merger bias parameters to their fiducial values and derive constraints for the remaining cosmological parameters, with a flat prior on them. In "run B" we consider the full set of parameters, including bias ones, and again take flat priors on all parameters, except $H_0$. In "run C", we set {\it Planck 2018} priors \cite{Planck_2018} on all cosmological parameters (to maximize constraining power in the study of merger bias). Summarizing: 
\vspace{-0.15cm}
\begin{itemize}[leftmargin=6mm]
    \item[\hspace{-0.2cm}] run A: $\Theta = [H_0,\Omega_ch^2,w_0,w_a]$, uniform prior on [$\Omega_ch^2,w_0,w_a$]; {\it Planck} prior on $H_0$;\vspace{-0.15cm}
    \item[\hspace{-0.2cm}] run B: $\Theta = [H_0,\Omega_ch^2,w_0,w_a,b_0...b_{n}]$,
    uniform prior on [$\Omega_ch^2,w_0,w_a$]; {\it Planck} prior on $H_0$;\vspace{-0.15cm}
    \item[\hspace{-0.2cm}] run C: $\Theta = [H_0,\Omega_ch^2,w_0,w_a,b_0...b_n]$, {\it Planck} prior on all the cosmological parameters. 
\end{itemize}

\subsection{Cosmological parameter constraints}\label{sec:cosmology_res}

In Tab. \ref{tab:res_cosmo}, we report the marginalized $1\sigma$ errors computed for each of the cosmological parameters $[\Omega_ch^2,w_0,w_a]$ separately for a DBH and a DNS survey. 
The ET-like results, with either the conservative or the optimistic $k_{nl}^0$ cut-offs, are reported in the first two rows of the Table, considering a localization error $\Delta\Omega = 3\ \text{deg}^2$ for DBH and $\Delta\Omega = 10\ \text{deg}^2$ for DNS, whereas the remaining entries consider more advanced scenarios.

For these, we consider three improvements, all of which computed in the conservative $k_{nl}^0$ case. 
First of all, in the third row, we improve the sky localization error to $\Delta \Omega = 0.5 \ \deg^2$.
Second of all, we analyze a situation in which distances and sky localizations are measured with higher precision than in the ET analysis (as in the futuristic case described in Sec. \ref{sec:bins}), but we keep the number of sources unchanged with respect to the ET-like case. Results for this case are reported in the fourth row.
Finally, the fifth row considers the same high precision configuration, but increases the number of the observed sources to $\simeq 10^7$ separately for DBH and DNS. In this last case, the higher density of observed mergers is modelled by rescaling the expression in eq. (\ref{eq:dNdzdOmega_skew}), in order to get a higher total number of sources.

\begin{figure}[t!]
\includegraphics[width = 0.99\textwidth, height = 0.43\textheight]{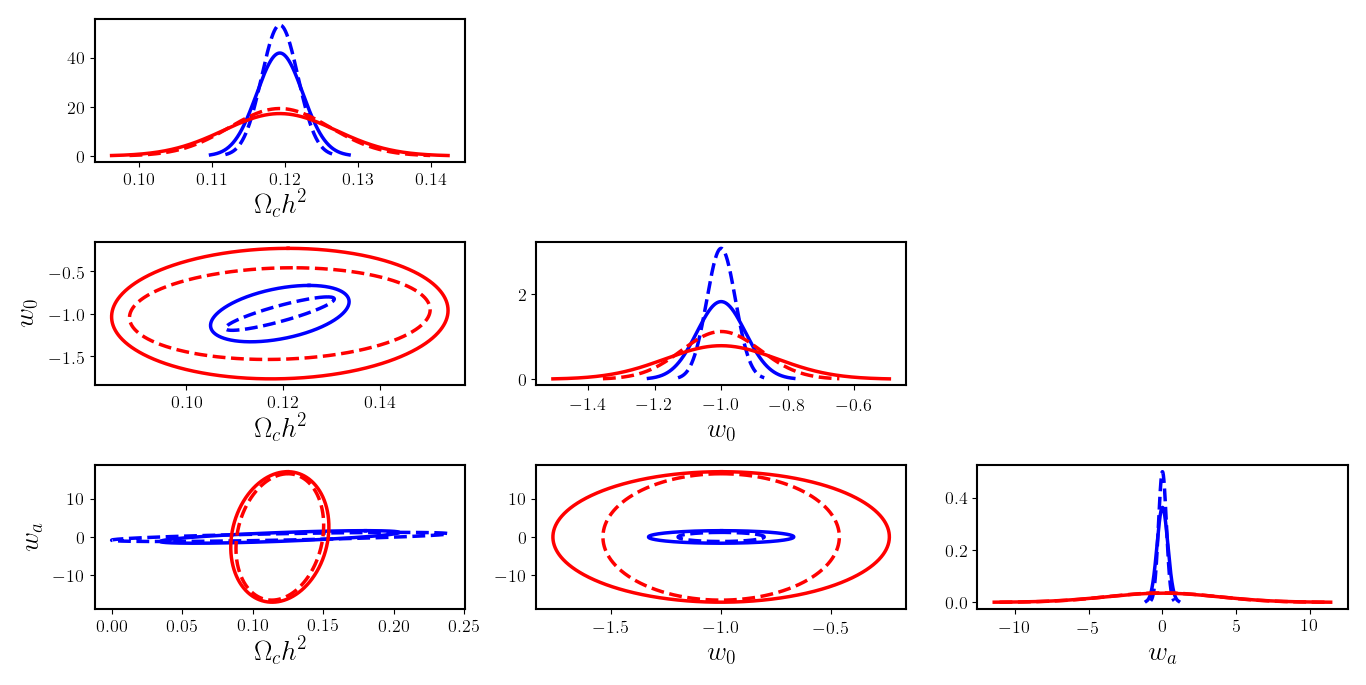}
\caption{Confidence $1\sigma$ ellipses obtained for DNS (red) and DBH (blue) in the ET-like survey run B, for each couple of cosmological parameters $(\Theta_{\alpha},\Theta_{\beta})$ described in Tab. \ref{tab:res_cosmo}. The plots for $(\Theta_{\alpha},\Theta_{\alpha})$ show the posterior distributions obtained. The dotted line shows the results obtained setting $k_{nl}^0=0.4\ h\text{Mpc}^{-1}$, while the continuous line refers to $k_{nl}^0=0.1\ h\text{Mpc}^{-1}$, both with $\Delta \Omega = 10 \ \text{deg}^2$ for DNS and $\Delta \Omega = 3 \ \text{deg}^2$ for DBH.}
\label{fig:ellipse}
\end{figure}

The run B results are used to compute the confidence ellipses in Fig. \ref{fig:ellipse}, which refer to the ET-like configuration. 
If $H_0$ was not marginalized, its forecasted $1\sigma$ error for the ET-like case assuming $k_{nl}^0= 0.4\ h\text{Mpc}^{-1}$, would be $7.0495$ for DBH and $17.374$ for DNS in run A, $16.180$ for DBH and $87.690$ for DNS in run B (both assuming uniform prior) and $1.5334$ for DBH and $1.5428$ for DNS in run C (assuming {\it Planck} prior \cite{Planck_2018}). 

\begin{table}[t!]
\caption{Forecasted $1\sigma$ marginalized errors for the cosmological parameters in the different studied scenarios for both DBH and DNS. The first two lines are our baseline case: they assume the ET specifications described in the main text, with different choices for $k_{nl}^0$. The third line has the same $D_L$ error as for the baseline case (i.e., same radial binning), but it assumes better sky localization.
In the fourth and fifth line, the $D_L$ measurement error is improved with respect to the baseline ET-configuration, while the number of sources is either kept at $N_{DBH}=10^{4.79},\ N_{DNS} =10^{4.14}$ or increased via a re-scaling of their distribution to reach $N_{DBH,DNS}=10^7$. In both these cases, the sky localization error is assumed to be negligible at all scales considered in the analysis (that corresponds to a localization with $\sim$ few arcmin$^2$ precision).
Different columns consider different choices of baseline parameters and priors for both DBH and DNS. Run A uses $\Theta = [H_0,\Omega_ch^2,w_0,w_a]$; run B and run C instead use $\Theta = [H_0,\Omega_ch^2,w_0,w_a,b_m^0...b_m^n]$. In all the cases, an $H_0$ {\it Planck} prior \cite{Planck_2018} is assumed. For all the other cosmological parameters, run A and B assume a uniform prior, whereas run C adopts {\it Planck} priors \cite{Planck_2018} (see Tab. \ref{tab:fiducial_cosmology}).}\vspace{0.3cm}
\label{tab:res_cosmo}
\centering
\begin{tabular}{|c|c||c|c|c||c|c|c|}
\toprule
$D_L$ error & Parameter & \multicolumn{3}{c||}{DBH}& \multicolumn{3}{c|}{DNS}\\
& & run A & run B & run C& run A & run B & run C\\
\midrule
\multirow{3}{*}{\shortstack[c]{Baseline $\Delta\Omega$ \\ $k_{nl}^0 = 0.1\ h\text{Mpc}^{-1}$}} &
$\Omega_ch^2$ & $0.0037$ & $0.0095$ & $0.0082$
& $0.0192$ & $0.0230$ & $0.0191$ \\
& $w_0$ & $0.1460$ & $0.2185$ & $0.1911$
& $0.4697$ & $0.5058$ & $0.4951$ \\
& $w_a$ & $0.5030$ & $1.0941$ & $0.8487$
& $1.3186$ & $11.378$ & $1.3390$ \\
\midrule
\multirow{3}{*}{\shortstack[c]{Baseline $\Delta\Omega$ \\ $k_{nl}^0 = 0.4\ h\text{Mpc}^{-1}$}} &
$\Omega_ch^2$ & $0.0025$ & $0.0075$ & $0.0068$
& $0.0165$ & $0.0206$ & $0.0168$ \\
& $w_0$ & $0.0797$ & $0.1296$ & $0.1205$
& $0.3239$ & $0.3554$ & $0.3525$ \\
& $w_a$ & $0.2993$ & $0.7946$ & $0.6843$
& $0.9026$ & $11.019$ & $1.3384$ \\
\midrule
\multirow{3}{*}{\shortstack[c]{$\Delta\Omega = 0.5\ \text{deg}^2$\\ $k_{nl}^0 = 0.1\ h\text{Mpc}^{-1}$}} &
$\Omega_ch^2$ & $0.0037$ & $0.0095$ & $0.0083$
& $0.0191$ & $0.0229$ & $0.0191$\\
& $w_0$ & $0.1453$ & $0.2177$ & $0.1906$
& $0.4615$ & $0.5063$ & $0.4953$ \\
& $w_a$ & $0.5009$ & $1.0951$ & $0.8491$
& $1.3191$ & $11.377$ & $1.3390$ \\
\midrule
\multirow{3}{*}{\shortstack[c]{"High precision" \\ $10^5$ sources\\ $k_{nl}^0 = 0.1\ h\text{Mpc}^{-1}$}} &
$\Omega_ch^2$ & $0.0033$ & $0.0084$ & $0.0076$
& $0.0050$ & $0.0090$ & $0.0083$ \\
& $w_0$ & $0.1250$ & $0.1675$ & $0.1536$
& $0.1423$ & $0.2001$ & $0.1848$ \\
& $w_a$ & $0.4319$ & $0.9710$ & $0.7875$
& $0.4587$ & $1.4172$ & $0.9765$ \\
\midrule
\multirow{3}{*}{\shortstack[c]{"High precision" \\ $10^7$ sources \\ $k_{nl}^0 = 0.1\ h\text{Mpc}^{-1}$}} &
$\Omega_ch^2$ & $0.0024$ & $0.0075$ & $0.0070$
& $0.0050$ & $0.0088$ & $0.0082$ \\
& $w_0$ & $0.0746$ & $0.1249$ & $0.1184$
& $0.1417$ & $0.1986$ & $0.1841$ \\
& $w_a$ & $0.2682$ & $0.8565$ & $0.7228$ 
& $0.4553$ & $1.3933$ & $0.9686$ \\
\bottomrule
\end{tabular}
\end{table}

For the ET-like forecasts in the strictly linear regime ($k_{nl}^0 = 0.1 \, h{\rm Mpc}^{-1}$), our results for both the merger kinds show that we can achieve error bars on $\Omega_ch^2$ and $w_0$ which are worse, but not far from those expected via galaxy clustering analysis in the near future (using for example the Euclid catalogue). For $w_a$ we instead find error bars which are about 5 times worse than those expected with Euclid in the DBH case, while they are almost 10 times worse in the case of DNS. This difference is due to the different redshift range covered by the two distributions (i.e. to the fact that DNS tracers can be used only up to $z^{max}\sim 2$). The same reason explains the difference in the $H_0$ forecasting. We verify that these expectations change only marginally (by a few percent) if we take a non-informative prior on $H_0$.
Of course, optimistically pushing the analysis into more non-linear scales significantly improves these figures. Likewise, much tighter constraints can be achieved with the improved settings (i.e. a higher precision in the sky position or distance determination, an higher number of sources, or all of them at once), compared to the baseline ET-case.

Regardless of its actual constraining power in different regimes, we argue anyway that the main interest of this type of analysis lies in the complementarity between merger and galaxy surveys. We have pointed out since the beginning that Gravitational Wave surveys are in Luminosity Distance Space and we have shown that Luminosity Distance Space Distortions behave differently from Redshift-Space Distortions. Moreover, GW surveys provide information up to very high redshift $z>2$. This means larger volumes than many forthcoming optical galaxy surveys. For scenarios with high precision sky localization, it also allows pushing the tomographic analysis up to small scales for high-$z$ shells, while staying in the linear or quasi-linear regime. Constraints on other interesting parameters, which we have not considered here, such as the primordial, local non-Gaussian ($f_{\rm NL}$) amplitude, generally significantly benefit from large survey volumes at high redshift ($z>2$), and the same time do not require small localization errors; we will investigate this further in the future. 

Another crucial opportunity offered by merger surveys, which we have already mentioned, is of course that of marginalizing over cosmological parameters and focusing instead on the study of merger bias. This is explored further in Sec. \ref{sec:bias_res}

\subsection{Bias parameter constraints}\label{sec:bias_res}
\begin{figure}[t!]
\centering
\subfigure{
\includegraphics[width = 0.9\textwidth]{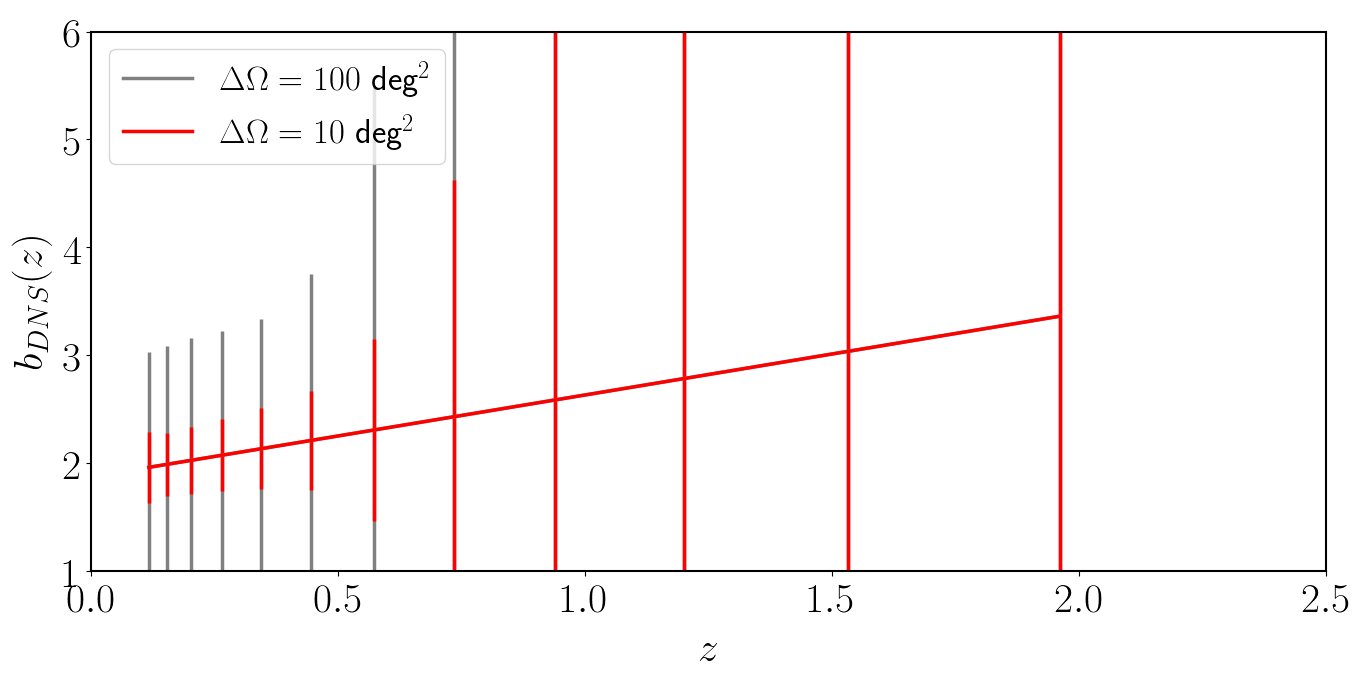}}
\subfigure{
\includegraphics[width = 0.9\textwidth]{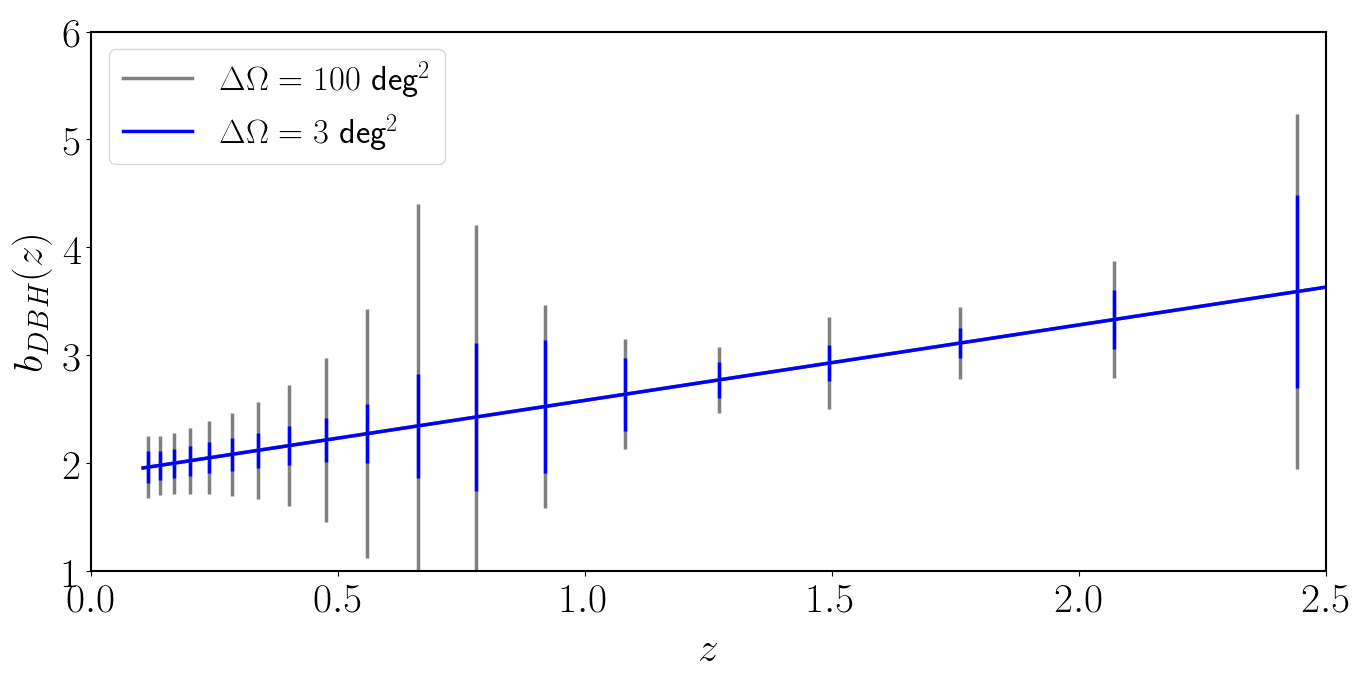}}
\caption{Fiducial bias with errors, forecasted through the Fisher matrix analysis in run C for the ET-like survey for DNS (on the top) and DBH (on the bottom) (see Tab. \ref{tab:res_bias}). Both the cases assume $k_{nl}^0 = 0.1\ h\text{Mpc}^{-1}$ and $\Delta\Omega = 100\ \text{deg}^2$ (grey line) or, respectively, $\Delta\Omega = 10\ \text{deg}^2$ for DNS (red line) and $\Delta\Omega = 3\ \text{deg}^2$ for DBH (blue line). Each point highlights the central $z_i$ of the bins in Tab. \ref{tab:DLbin}. For sake of clarity in showing the DBH results, only the even bins (i.e. $z_i$ where $i$ is even) for which $[\sigma_{b_{DBH}}/b_{DBH}](z_i)< 0.7$ are showed, that is $z<2.88$. For DNS, all the bins have been plotted instead. Having set $z^{max} = 2$, DNS bias model is cut after this value.}
\label{fig:bias_error}
\end{figure}

\begin{figure}[t!]
\centering
\subfigure{
\includegraphics[width = 0.9\textwidth]{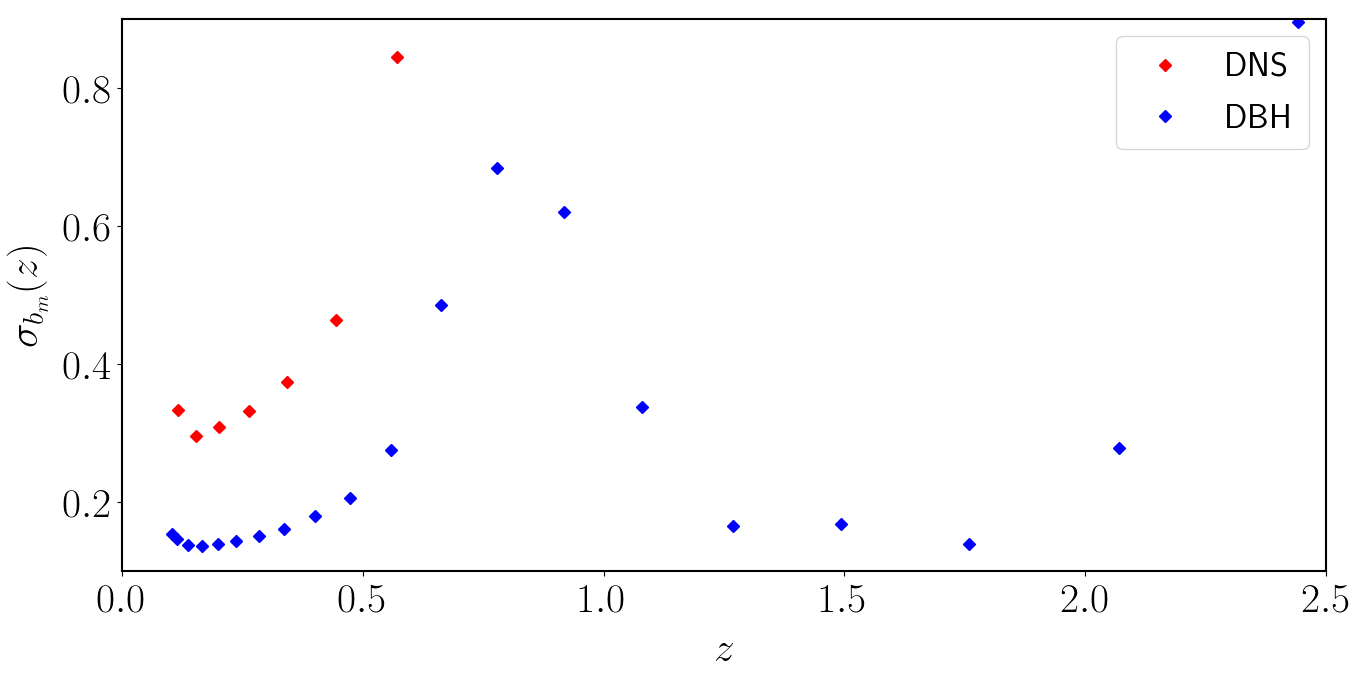}}
\subfigure{\includegraphics[width = 0.9\textwidth]{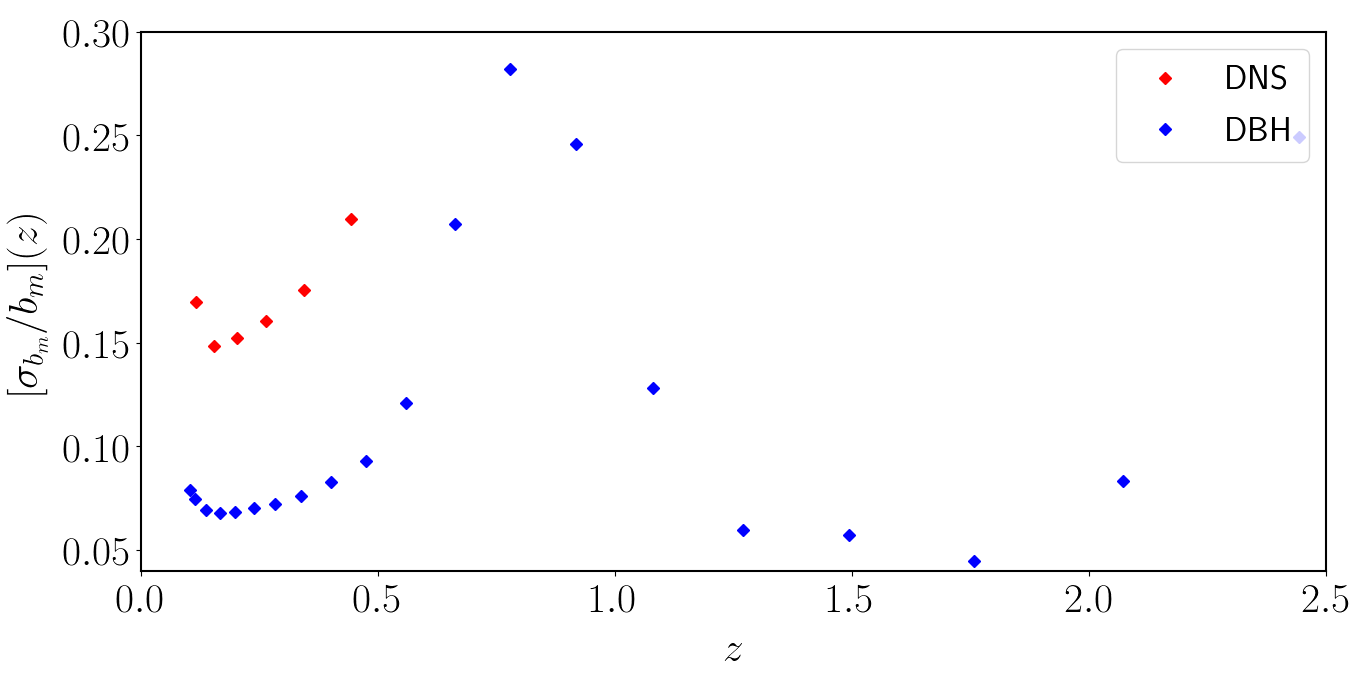}}
\caption{Bias errors in run C in the baseline case for both DNS (red, $\Delta\Omega = 10\ \text{deg}^2$) and DBH (blue, $\Delta\Omega = 3\ \text{deg}^2$) with $k_{nl}^0=0.1\ h\text{Mpc}^{-1}$. In the upper panel, the dots represent the errors $\sigma_{b_m}$ in the bins in Tab. \ref{tab:DLbin}. The lower panel shows instead the relative error, $[\sigma_{b_m}/b_m](z)$. For both DNS and DBH, the points refer to the same bins showed in Fig. \ref{fig:bias_error}.}
\label{fig:interp_sigma_bias}
\end{figure}

We focus now on merger bias parameters. Since the focus is on merger properties here, rather than on Cosmology, it is appropriate and useful to include stringent cosmological priors from e.g. CMB surveys such as {\it Planck}.
Therefore, in Appendix \ref{app:APS} in Tab. \ref{tab:res_bias} and \ref{tab:res_bias_lin}, we focus on results from run C ({\it Planck} cosmological priors \cite{Planck_2018}); they are obtained considering in the first case $\Delta \Omega = 3 \ \text{deg}^2$ for DBH and $\Delta\Omega = 10 \ \text{deg}^2$ for DNS, while in the second $\Delta \Omega = 100 \ \text{deg}^2$ for both the mergers. All the results assume $k_{nl}^0 = 0.1\ h\text{Mpc}^{-1}$. The case $k_{nl}^0 = 0.4\ h\text{Mpc}^{-1}$ does not provide significant improvements in the bias forecasts. 
Fig. \ref{fig:bias_error} shows both the fiducial values and error bars, $b_m(z_i) \pm \sigma_{b_m}(z_i)$, obtained in run C, adopting the model described in eq. (\ref{eq:bias_m}). Results are showed assuming either $\Delta \Omega = 3\ \text{deg}^2$ for DBH and $10 \ \text{deg}^2$ for DNS, or $\Delta \Omega = 100 \ \text{deg}^2$ for both the merger kinds. The modulation that can be seen in the error bars for the DBH bias (i.e. $\sigma_{b_{DBH}}(z_i = 1.5) < \sigma_{b_{DBH}}(z_i = 0.7)$) is due to a combination of two effects: the presence of a peak in the number of sources around $z\sim 1$ (compare Fig. \ref{fig:bias_error} with Fig. \ref{fig:dndzdomega}) on one side, and the increasing luminosity distance error on the other; this leads to a minimum in the error bar at $z \sim 1.2$. The same effect is not seen in the DNS case since the larger amplitude of the $D_L$ intervals (due to $\Delta D_L/D_L$) covers the modulation.
Note that each bias parameter refers to one of the $D_L$ bins: its fiducial value is computed using eq. (\ref{eq:bias_m}) in the central point $z_i$. 
The absolute and relative bias errors are displayed in Fig. \ref{fig:interp_sigma_bias}. We conclude that merger bias should be detected by ET at high significance, all the way up to $z \sim 2$ for DBH, up to $z \sim 0.5$ for DNS.

To verify whether our method also allows constraining merger bias without any cosmological assumption, we consider a new configuration, in which uniform priors are assumed on {\em all} the cosmological parameters (i.e., with a flat prior now also on $H_0$). Tab. \ref{tab:res_bias} and Tab. \ref{tab:res_bias_lin} report also the results for this "run D". Run C and run D are compared in Appendix \ref{app:APS} in Fig. \ref{fig:bias_no_cosmo}, for both DBH and DNS in the case of an ET-like survey with $k_{nl}^0=0.1\ h \text{Mpc}^{-1}$. The results of the C and D runs differs slightly at low $z$, particularly for the DNS case, while at high $z$ the difference between the two becomes negligible. Therefore, in this setting, it turns out that the constraining power on bias almost does not depend on the prior assumed for cosmology, particularly in the DBH case. 

Having this kind of measurement opens up new interesting prospects. For example, it would be interesting to consider different types of GW sources separately, to understand whether their different bias models could be distinguished from one another. One possible application for this involves the study of Primordial Black Holes (PBH). Since PBH form before galaxies, their bias -- and therefore the bias of their mergers -- should be different from those calculated in Sec. \ref{sec:merbias}. Therefore, the study of the APS from future GW surveys, particularly to understand their dependence on bias parameters, could help shedding light on the existence of PBH or on the properties of their distribution. Even in a more standard scenario, simply comparing the actual measured bias of compact binary mergers to predictions from theory and simulations would clearly already be of interest. 
We plan to analyze more ideas and applications of this kind in future studies.

\section{Conclusions}\label{sec:conclusions}

In this work, we have discussed the possibility of using clustering properties of GW from mergers in Luminosity Distance Space as a tool for Cosmology. This approach -- originally proposed in \cite{Namikawa_2016, Pengjie_2018} -- will be possible with networks of second and third generation detectors, which will detect $\sim 10^5$, or more, merger events, with good distance and sky localization precision. 
In our study, we have mostly focused on tomographic measurements of the Angular Power Spectrum of the DBH and DNS mergers for a network of three ET-like detectors. We have also considered more futuristic configurations, allowing for a higher number of sources and for a better distance and sky localization measurement, with respect to our ET-like baseline. We have produced Fisher matrix forecasts both for cosmological parameters (matter and dark energy) and for the bias parameters of both kinds of mergers, forecasting the latter with and without priors from external data sets (e.g., constraints on Cosmology from {\it Planck}).
We have concentrated our attention on DBH and DNS mergers and we have predicted both their fiducial number densities and bias, starting from the astrophysically motivated results of \cite{Artale_2018, Artale_2020}, which combine hydrodinamical cosmological simulations with population synthesis models. Our bias models were built using a Halo Occupation Distribution approach.  

Our final expected constraints on cosmological parameters are less powerful than those achievable in the near future via galaxy clustering studies. This was essentially foreseeable, in light of the smaller number of tracers which we expect for ET, compared to, e.g., Euclid. It must however be noticed that the large volumes and high redshifts probed with GW mergers, particularly with DBH, partially compensate for this issue, and still lead to interesting results for some parameters, such as, e.g., $\Omega_c h^2$ or $w_0$.
Of course, if we instead consider experiments that will detect a larger number of events than ET, with high precision determination of distances and sky localization -- or a longer observation time for ET itself -- the cosmological forecasts significantly improve and lead to potentially tight constraints.

Regardless of the exact expected constraints and of the survey under study, the main point of interest of this approach relies anyway in the complementarity between GW and galaxy survey analyses. As already mentioned above, compact binary mergers detected by ET will extend up to very high redshifts, where electromagnetic counterparts are not available. Furthermore, Luminosity Distance Space Distortions -- which have to be considered in the GW analysis -- have a different structure with respect to Redshift-Space Distortions in galaxy catalogues. Another crucial aspect is that this method does not rely on any external dataset or assumption, since it does not require to infer the redshift of GW events.  

Finally, besides focusing on cosmological parameters, we have also explicitly shown how the approach investigated in this work will allow us to measure the bias of DBH and DNS mergers at high statistical significance, over a large redshift range (we find that this is possible also with much lower precision in sky localization, with respect to that required to achieve meaningful cosmological parameter constraints). This in turn can provide interesting information about the physical nature and properties of mergers themselves. This, and other interesting applications will be the object of further investigation in the future.

\newpage
\appendix

\section{Simulations}\label{app:sim}
The catalogs of binary compact object mergers adopted here come from \cite{Artale_2018,Artale_2020}. These were obtained by seeding the galaxies from the {\sc eagle} cosmological simulation \citep{Schaye2015} with binary compact objects from population-synthesis simulations \citep{Giacobbo2018b,Mapelli2018}.
The {\sc eagle} suite is a set of cosmological simulations \citep{Schaye2015} run with a modified version of the smoothed particle hydrodynamics {\sc gadget-3}, that  tracks the evolution of gas, dark matter and stars across cosmic time in a simulated Universe.
The simulation includes sub-grid models for cooling, star formation, chemical enrichment, stellar and active galactic nucleus feedback. The parameters of the sub-grid models are constrained to match observational results of the stellar mass function, and stellar mass-black hole mass relation at $z=0$. In this work we use the binary compact object mergers from the galaxy catalog of the highest resolution box of (25 Mpc)$^3$.
Binary compact objects are randomly associated with stellar particles in the cosmological simulation based on the formation time, metallicity and total initial mass of each stellar particle \citep{Mapelli2017}. Thanks to this algorithm, we self-consistently take into account the properties of the stellar progenitors of each binary compact object, as well as the delay time between formation and merger of the binary.
The initial population-synthesis simulations were run with {\sc mobse} \citep{Giacobbo2018}. {\sc mobse} exploits: 1) fitting formulas to describe the evolution of stellar properties as a function of metallicity and stellar mass (e.g. radii and luminosity, \cite{Hurley2000}); 2) up-to-date models for stellar-wind mass loss \citep{Giacobbo2018}; 3) state-of-the-art prescriptions for core-collapse \citep{Fryer2012} and pair-instability supernovae \citep{Mapelli2020}; 4) a formalism for binary-evolution processes \citep{Hurley2002}. The mass function and local merger rate density obtained with {\sc mobse} are in agreement with results from O1, O2 and O3 of Advanced LIGO and Virgo \citep{LVC_3,LVC_4,O3}. We refer to \cite{Giacobbo2018b} and \citep{Artale_2020} for more detail on population-synthesis and cosmological simulations, respectively.
\vspace{0.1cm}
\subsection{Simulation snapshots}
\vspace{-0.6cm}
\begin{table}[ht!]
\caption{Redshift and time snapshots ($z = 2.22\cdot 10^{-16}$, $z = 0.1$ are considered together in Sec. \ref{sec:sources}). }\vspace{0.3cm}
\label{tab:snap}
\centering
\begin{tabular}{|c|c||c|c||c|c||c|c|}
\toprule
$z$ & $T^{SIM} [\text{Gyr}]$ & $z$ & $T^{SIM} [\text{Gyr}]$ & $z$ & $T^{SIM} [\text{Gyr}]$& $z$ & $T^{SIM} [\text{Gyr}]$ \\
\midrule
2.22 $\cdot 10^{-16}$ & 0.676 & 
0.61 & 0.737 &
1.74 & 0.525 &
3.98 & 0.223
\\
0.10 & 1.161 &
0.73 & 0.685 &
2.00 & 0.409 &
4.49 & 0.194 
\\
0.18 & 0.947 &
0.86 & 0.634 &
2.24 & 0.312 &
5.04 & 0.150 
\\
0.27 & 0.902 &
1.00 & 0.757 &
2.48 & 0.402 &  
5.49 & 0.113 
\\
0.37 & 0.987 &
1.26 & 0.770 &
3.02 & 0.429 & 
6.00 & 0.145 
\\
0.50 & 0.930 &
1.49 & 0.596 &
3.53 & 0.294 &
&
\\
\bottomrule
\end{tabular}
\end{table}

\vspace{-0.2cm}
\subsection{Stellar mass and star formation rate bins}
\vspace{-0.6cm}
\begin{table}[hb!]
\caption{Stellar mass ($M_{\odot}$ units) and star formation rate ($M_{\odot}yr^{-1}$ units) bins for the host galaxies.}\vspace{0.3cm}
\label{tab:stellar_mass}
\centering
\begin{tabular}{|c|c|c||c|c|c|}
\toprule
$\log M_*$ bins & $\log M_*$ bins & $\log M_*$ bins &  $\log SFR$ bins  &  $\log SFR$ bins &  $\log SFR$ bins \\
\midrule
$7.00,7.33$ & $7.33,7.67$  & $7.67,8.00$ & $-5.50,-4.97$ & $-4.97,-4,43$  & $-4.43,-3.90$ \\
$8.00,8.33$ & $8.33,8.67$ & $8.67,9.00$ & $-3.90,-3.37$ & $-3.37,-2.83$ & $-2.83,-2.30$\\
$9.00,9.33$ & $9.33,9.67$  &  $9.67,10.0$ & $-2.30,-1.77$ & $-1.77,-1.23$  & $-1.23,-0.70$ \\
$10.0,10.3$ & $10.3,10.7$ & $10.7,11.0$ & $-0.70,-0.17$ & $-0.17, 0.37$ & $ 0.37, 0.90$ \\
$11.0,11.3$ & $11.3,11.7$  & $11.7,12.0$  & $ 0.90, 1.43$ & $ 1.43, 1.97$ & $ 1.97, 2.50$  \\
\bottomrule
\end{tabular}
\end{table}

\clearpage
\section{Angular Power Spectrum and Fisher computation}\label{app:APS}
This appendix reports information on the setting used to compute the APS and the forecasts.

\subsection{Survey setting}
\vspace{-0.6cm}
\begin{table}[ht!]
\centering
\caption{Survey specifications that were assumed in our forecasts. In all cases we assume full sky coverage ($f_{sky} = 1$) and we extend the analysis up to redshift $z^{max} = 5$ for DBH, $z^{max} = 2$ for DNS. Details can be found in Sec. \ref{sec:LDSD_implement}, \ref{sec:APS} and \ref{sec:fisher}.}\vspace{0.3cm}
\label{tab:survey_details}
\centering
\begin{tabular}{|c||c|}
\toprule
\multicolumn{2}{|c|}{Survey setting}  \\
\midrule
\multirow{3}{*}{{\shortstack[c]{$3$yr ET-like survey \\ 3-detector network}}} & Sources: $\sim 10^5$ \\
&  {$\Delta D_L / D_L = 10\%$ for DBH, $30\%$ for DNS}\\
 & {$\Delta \Omega = 3\ \text{deg}^2$ for DBH, $10\ \text{deg}^2$ for DNS} \\
\midrule
\multirow{3}{*}{{\shortstack[c]{$3$yr ET-like survey \\ 3-detector network}}} & Sources: $\sim 10^5$ \\
&  {$\Delta D_L / D_L = 10\%$ for DBH, $30\%$ for DNS}\\
& $\Delta \Omega = 0.5\ \text{deg}^2$ \\
\midrule
\multirow{3}{*}{{\shortstack[c]{"High precision" \\ case 1}}} & Sources: $\sim 10^5$ \\
 & $\Delta D_L / D_L = 10\% \ \text{if}\ z < 2, 3\% \ \text{if}\ z \geq 2$\\  
 &  $\Delta \Omega \sim \text{few arcmin}^2$ \\
\midrule
\multirow{3}{*}{{\shortstack[c]{"High precision" \\ case 2}}}  & Sources: $\sim 10^7$ \\
 & $\Delta D_L / D_L = 10\% \ \text{if}\ z < 2, 3\% \ \text{if}\ z \geq 2$\\ 
 &  $\Delta \Omega \sim \text{few arcmin}^2$ \\
\bottomrule
\end{tabular}
\end{table}

\vspace{-0.2cm}
\subsection{Fiducial Cosmology}
\vspace{-0.6cm}
\begin{table}[ht!]
\caption{Fiducial values of the cosmological parameters from {\it Planck 2018} \cite{Planck_2018}. The ones which are associated with an error ($68\%$ limit) are the ones used in Sec. \ref{sec:fisher} to compute the Fisher matrix. The values in the first, third and fourth lines are taken from TT,TE,EE+lowE+lensing+BAO data. The values in the second line are compatible with the $\Lambda$CDM model in a spatially flat Universe; in particular, the errors for the DE EoS parameters $w_0$ and $w_a$ have been estimated from the ones in Planck+SNe+BAO data. The magnitude bias $s$ is set through $5s-2 = 0$ to have no magnification.}\vspace{0.3cm}
\label{tab:fiducial_cosmology}
\centering
\begin{tabular}{|c||c||c|}
\toprule
\multicolumn{3}{|c|}{Fiducial Cosmology}  \\
\midrule
$H_0 = 67.66\pm 0.42$ & $\Omega_ch^2 = 0.11933 \pm 0.00091$ & $\Omega_bh^2 = 0.02242$ \\
$\Omega_k = 0.0$ & $w_0 = -1\pm 0.13$ & $w_a  = 0\pm 0.55$ \\
$A = 2.105\cdot 10^{-9}$ & $n_s = 0.9665$ & $T_{CMB} = 2.7255$\\
$N_{eff} = 2.99$  & $ Y_{He} = 0.242$ & $\tau = 0.0561$ \\
$\Delta z_{rei} = 0.5$ & $\Omega_{\nu}h^2 = 0.00064$ & $s = 0.4$ \\
\bottomrule
\end{tabular}
\end{table}

\newpage
\subsection{Distance bins}
\vspace{-0.2cm}
\LTcapwidth=\textwidth
\begin{longtable}{|c|c||c|c||c|c|}
\caption{{\small Luminosity distance bins in the ET-like survey respectively for DBH and DNS. The associated central redshifts $z_i$ are computed in the fiducial Cosmology (Tab. \ref{tab:fiducial_cosmology}). In the DBH case, $47$ bins are obtained setting $\Delta D_L/D_L = 10\%$, $z^{min} = 0.1$, $z^{max} \simeq 5$; in the DNS case, $12$ bins are obtained setting $\Delta D_L/D_L = 30\%$, $z^{min} = 0.1$, $z^{max} \simeq 2$. For details see Sec. \ref{sec:APS}.}}
\label{tab:DLbin}\\
\hline
\multicolumn{6}{|c|}{DBH} \\
\hline
$D_L$ bins [Mpc] & $\quad z_i \quad $ & 
$D_L$ bins [Mpc] & $\quad z_i \quad $ & 
$D_L$ bins [Mpc] & $\quad z_i \quad $  \\
\hline
\endfirsthead
\multicolumn{6}{c}{\footnotesize{Continued from previous page}} \\
$D_L$ bins [Mpc] & $\quad z_i \quad $  & 
$D_L$ bins [Mpc] & $\quad z_i \quad $  & 
$D_L$ bins [Mpc] & $\quad z_i\quad $  \\
\hline
\endhead
\multicolumn{6}{c}{\footnotesize{Continued on next page}} \\
\endfoot
\hline
\endlastfoot
$475.73, 525.81$ & $0.10$ &
$525.81, 581.16$ & $0.12$ &
$581.16, 642.33$ & $0.13$ \\
$642.33, 709.94$ & $0.14$ &
$709.94, 784.67$ & $0.15$ & 
$784.67, 867.27$ & $0.17$  \\
$867.27, 958.56$ & $0.18$  &
$958.56, 1059.46$ & $0.20$ &
$1059.46, 1170.99$ & $0.22$ \\
$1170.99, 1294.25$ & $0.24$  &
$1294.25, 1430.49$ & $0.26$  &
$1430.49, 1581.06$ & $0.28$  \\
$1581.06, 1747.49$ & $0.31$ &
$1747.49, 1931.44$ & $0.34$  &
$1931.44, 2134.75$ & $0.37$  \\
$2134.75, 2359.46$ & $0.40$  &
$2359.46, 2607.82$ & $0.44$  &
$2607.82, 2882.33$ & $0.47$  \\
$2882.33, 3185.73$ & $0.51$  & 
$3185.73, 3521.07$ & $0.56$  &
$3521.07, 3891.71$ & $0.61$   \\ 
$3891.71, 4301.36$ & $0.66$  &
$4301.36, 4754.14$ & $0.72$  & 
$4754.14, 5254.57$ & $0.78$ \\
$5254.57, 5807.68$ & $0.85$  &
$5807.68, 6419.02$ & $0.92$ &
$6419.02, 7094.71$ & $1.00$  \\ 
$7094.71, 7841.52$ & $1.08$  &
$7841.52, 8666.94$ & $1.17$  & 
$8666.94, 9579.25$ & $1.27$  \\
$9579.25, 10587.6$ & $1.38$ &
$10587.6, 11702.1$ & $1.49$ &
$11702.1, 12933.9$ & $1.62$  \\
$12933.9, 14295.3$ & $1.76$  &
$14295.3, 15800.1$ & $1.91$  &
$15800.1, 17463.3$ & $2.07$  \\
$17463.3, 19301.5$ & $2.25$  &
$19301.5, 21333.2$ & $2.44$  &
$21333.2, 23578.9$ & $2.65$ \\
$23578.9, 26060.8$ & $2.88$ &
$26060.8, 28804.1$ & $3.13$  &
$28804.1, 31836.1$ & $3.41$  \\
$31836.1, 35187.3$ & $3.70$  &
$35187.3, 38891.2$ & $4.03$ &
$38891.2, 42985.0$ & $4.39$  \\
$42985.0, 47509.7$ & $4.78$  &
$47509.7, 52510.8$ & $5.20$  &
&  \\
\hline
\multicolumn{6}{|c|}{DNS} \\
\hline
$D_L$ bins [Mpc] & $\quad z_i \quad $  & 
$D_L$ bins [Mpc] & $\quad z_i \quad $  & 
$D_L$ bins [Mpc] & $\quad z_i\quad $  \\
\hline
$475.73,643.6$ & $0.12$ & 
$643.63,870.8$ & $0.15$ &
$870.80,1178.1$ & $0.20$\\
$1178.14,1593.96$ & $0.26$ & 
$1593.96,2156.53$ & $0.34$ & 
$2156.53,2917.66$ & $0.44$ \\
$2917.66,3947.42$ & $0.57$ &
$3947.42,5340.62$ & $0.73$ & 
$5340.62,7225.55$ & $0.94$ \\
$7225.55,9775.74$ & $1.20$ & 
$9775.74,13226.0$ & $1.53$ & 
$13226.0,17894.0$ & $1.96$\\
\end{longtable}

\LTcapwidth=\textwidth
\begin{longtable}{|c|c||c|c||c|c|}
\caption{{\small Luminosity distance bins in which the APS is computed for the futuristic "high precision" configuration separately for DBH and DNS. The associated central redshifts $z_i$ are computed in the fiducial cosmology (Tab. \ref{tab:fiducial_cosmology}). The $70$ (DBH) and $36$ (DNS) bins are obtained setting $\Delta D_L/D_L = 10\%$ if $z < 2$ and $3\%$ if $z \geq 2$ , $z^{min} = 0.1$, $z^{max} \simeq 5$ (DBH) or $z^{max} \simeq 2$ (DNS) (see Sec. \ref{sec:bins}). For details see Sec. \ref{sec:APS}.}}
\label{tab:DLbin_DECIGO}\\
\hline
\multicolumn{6}{|c|}{DBH} \\
\hline
$D_L$ bins [Mpc] & $\quad z_i\quad $ & $D_L$ bins [Mpc] & $\quad z_i\quad$ & $D_L$ bins [Mpc] & $\quad z_i\quad$ \\
\hline
\endfirsthead
\multicolumn{6}{c}{\footnotesize{Continued from previous page}} \\
$D_L$ bins [Mpc] & $\quad z_i \quad $ & $D_L$ bins [Mpc] & $\quad z_i \quad $ & $D_L$ bins [Mpc] & $\quad z_i \quad $  \\
\hline
\endhead
\multicolumn{6}{c}{\footnotesize{Continued on next page}} \\
\endfoot
\hline
\endlastfoot
$475.73, 525.81$ & $0.10$ & 
$525.81, 581.16$ & $0.12$ &
$581.16, 642.33$ & $0.13$  \\ 
$642.33, 709.94$ & $0.14$  &
$709.94, 784.67$ & $0.15$ & 
$784.67, 867.27$ & $0.17$  \\
$867.27, 958.56$ & $0.18$ & 
$958.56, 1059.46$ & $0.20$ &
$1059.46, 1170.99$ & $0.22$ \\ 
$1170.99, 1294.25$ & $0.24$  & 
$1294.25, 1430.49$ & $0.26$  & 
$1430.49, 1581.06$ & $0.28$ \\
$1581.06, 1747.49$ & $0.31$ & 
$1747.49, 1931.44$ & $0.34$  &
$1931.44, 2134.75$ & $0.37$  \\ 
$2134.75, 2359.46$ & $0.40$  &
$2359.46, 2607.82$ & $0.44$   & 
$2607.82, 2882.33$ & $0.47$  \\ 
$2882.33, 3185.73$ & $0.51$  & 
$3185.73, 3521.07$ & $0.56$  &
$3521.07, 3891.71$ & $0.61$  \\ 
$3891.71, 4301.36$ & $0.66$  &
$4301.36, 4754.14$ & $0.72$  & 
$4754.14, 5254.57$ & $0.78$  \\ 
$5254.57, 5807.68$ & $0.85$  &
$5807.68, 6419.02$ & $0.92$ & 
$6419.02, 7094.71$ & $1.00$ \\ 
$7094.71, 7841.52$ & $1.08$  &
$7841.52, 8666.94$ & $1.17$  & 
$8666.94, 9579.25$ & $1.27$  \\
$9579.25, 10587.6$ & $1.38$ & 
$10587.6, 11702.1$ & $1.49$ & 
$11702.1, 12933.9$ & $1.62$  \\ 
$12933.9, 14295.3$ & $1.76$ &
$14295.3, 15800.1$ & $1.91$  & 
$15800.1, 17463.3$ & $2.07$ \\
$17463.3, 17995.1$ & $2.18$ & 
$17995.1, 18543.2$ & $2.24$  &
$18543.2, 19108.0$ & $2.29$ \\
$19108.0, 19690.0$ & $2.35$  &
$19690.0, 20289.7$ & $2.41$ & 
$20289.7, 20907.6$ & $2.47$  \\
$20907.6, 21544.4$ & $2.53$ & 
$21544.4, 22200.6$ & $2.59$ &
$22200.6, 22876.7$ & $2.66$  \\ 
$22876.7, 23573.5$ & $2.73$  & 
$23573.5, 24291.5$ & $2.80$  & 
$24291.5, 25031.3$ & $2.87$  \\
$25031.3, 25793.7$ & $2.94$  & 
$25793.7, 26579.3$ & $3.01$ & 
$26579.3, 27388.8$ & $3.09$ \\ 
$27388.8, 28223.0$ & $3.17$ &
$28223.0, 29082.5$ & $3.25$ & 
$29082.5, 29968.3$ & $3.33$ \\
$29968.3, 30881.0$ & $3.42$ & 
$30881.0, 31821.6$ & $3.50$ &
$31821.6, 32790.8$ & $3.59$ \\ 
$32790.8, 33789.5$ & $3.68$ & 
$33789.5, 34818.6$ & $3.78$  & 
$34818.6, 35879.1$ & $3.87$  \\
$35879.1, 36971.8$ & $3.97$ & 
$36971.8, 38097.9$ & $4.08$ & 
$38097.9, 39258.2$ & $4.18$ \\ 
$39258.2, 40453.9$ & $4.29$ &
$40453.9, 41686.0$ & $4.40$ & 
$41686.0, 42955.6$ & $4.51$  \\ 
$42955.6, 44263.9$ & $4.63$ & 
$44263.9, 45612.1$ & $4.75$ & 
$45612.1, 47001.3$ & $4.87$ \\ 
$47001.3, 48432.8$ & $5.00$ &
&&&\\
\hline
\multicolumn{6}{|c|}{DNS} \\
\hline
$D_L$ bins [Mpc] & $\quad z_i \quad $ & $D_L$ bins [Mpc] & $\quad z_i \quad $ & $D_L$ bins [Mpc] & $\quad z_i \quad $  \\
\hline
$475.73,525.81$ & $0.10$ &
$525.81,581.16$ & $0.12$ &
$581.16,642.33$ & $0.13$\\
$642.33,709.94$ & $0.14$ & 
$709.94,784.67$ & $0.15$ &
$784.67,867.27$ & $0.17$\\
$867.27,958.56$ & $0.18$ &
$958.56,1059.46$ & $0.20$ &
$1059.46,1170.99$ & $0.22$\\
$1170.99,1294.25$ & $0.24$ & 
$1294.25,1430.49$ & $0.26$ &
$1430.49,1581.06$ & $0.28$\\
$1581.06,1747.49$ & $0.31$ & 
$1747.49,1931.44$ & $0.34$ & 
$1931.44,2134.75$ & $0.37$\\
$2134.75,2359.46$ & $0.40$ & 
$2359.46,2607.82$ & $0.44$ & 
$2607.82,2882.33$ & $0.47$\\
$2882.33,3185.73$ & $0.51$ & 
$3185.73,3521.07$ & $0.56$ &
$3521.07,3891.71$ & $0.61$ \\
$3891.71,4301.36$ & $0.66$ &
$4301.36,4754.14$ & $0.72$ &
$4754.14,5254.57$ & $0.78$ \\
$5254.57,5807.68$ & $0.85$ &
$5807.68,6419.02$ & $0.92$ &
$6419.02,7094.71$ & $1.00$ \\
$7094.71,7841.52$ & $1.08$ &
$7841.52,8666.94$ & $1.17$ &
$8666.94,9579.25$ & $1.27$ \\
$9579.25,10587.59$ & $1.38$ &
$10587.6,11702.08$ & $1.49$ &
$11702.1,12933.87$ & $1.62$\\
$12933.9,14295.33$ & $1.76$ &
$14295.3,15800.11$ & $1.91$ &
$15800.1,17463.27$ & $2.07$\\
\end{longtable}

\vspace{0.2cm}
\subsection{Bias errors}
\vspace{-0.2cm}
\LTcapwidth=\textwidth
\begin{longtable}{|c|c|c|c|c|c|}
\caption{{\small ET-like survey with $k_{nl}^0 = 0.1\ h\text{Mpc}^{-1}$ for DBH and DNS. For each bin from Tab. \ref{tab:DLbin}, the central $z_i$ is indicated, together with the fiducial bias from eq. (\ref{eq:bias_m}). $1\sigma$ marginalized errors and relative errors $[\sigma_{b_m}/b_m](z_i)$ are shown both for run C and run D (see Sec. \ref{sec:bias_res}).}}
\label{tab:res_bias}\\
\hline
\multicolumn{6}{|c|}{DBH} \\
\hline
$z_i$& $b_m(z_i)$ & $\sigma_{b_m}$ run C & $\sigma_{b_m}$ run D & $[\sigma_{b_m}/b_m](z_i)$ run C & $[\sigma_{b_m}/b_m](z_i)$ run D \\
\hline
\endfirsthead
\multicolumn{6}{c}{\footnotesize{Continued from previous page}} \\
$z_i$& $b_m(z_i)$ & $\sigma_{b_m}$ run C & $\sigma_{b_m}$ run D & $[\sigma_{b_m}/b_m](z_i)$ run C & $[\sigma_{b_m}/b_m](z_i)$ run D \\
\hline
\endhead
\multicolumn{6}{c}{\footnotesize{Continued on next page}} \\
\endfoot
\hline
\endlastfoot
$0.10$ & $1.9534$ & $0.1537$ & $0.2086$ & $0.0787$ & $0.1068$\\
$0.12$ & $1.9606$ & $0.1459$ & $0.2026$ & $0.0744$ & $0.1033$\\
$0.13$ & $1.9685$ & $0.1400$ & $0.1969$ & $0.0711$ & $0.1000$\\
$0.14$ & $1.9770$ & $0.1371$ & $0.1919$ & $0.0694$ & $0.0971$\\
$0.15$ & $1.9863$ & $0.1359$ & $0.1870$ & $0.0684$ & $0.0941$\\
$0.17$ & $1.9965$ & $0.1355$ & $0.1822$ & $0.0679$ & $0.0913$\\
$0.18$ & $2.0076$ & $0.1363$ & $0.1784$ & $0.0679$ & $0.0889$\\
$0.20$ & $2.0196$ & $0.1383$ & $0.1758$ & $0.0685$ & $0.0871$\\
$0.22$ & $2.0326$ & $0.1406$ & $0.1738$ & $0.0692$ & $0.0855$\\
$0.24$ & $2.0468$ & $0.1435$ & $0.1721$ & $0.0701$ & $0.0841$\\
$0.26$ & $2.0622$ & $0.1465$ & $0.1704$ & $0.0710$ & $0.0826$\\
$0.28$ & $2.0789$ & $0.1502$ & $0.1696$ & $0.0723$ & $0.0816$\\
$0.31$ & $2.0969$ & $0.1550$ & $0.1701$ & $0.0739$ & $0.0811$\\
$0.34$ & $2.1165$ & $0.1610$ & $0.1728$ & $0.0761$ & $0.0817$\\
$0.37$ & $2.1377$ & $0.1689$ & $0.1785$ & $0.0790$ & $0.0835$\\
$0.40$ & $2.1606$ & $0.1793$ & $0.1878$ & $0.0830$ & $0.0869$\\
$0.44$ & $2.1854$ & $0.1926$ & $0.2006$ & $0.0881$ & $0.0918$\\
$0.47$ & $2.2113$ & $0.2059$ & $0.2154$ & $0.0931$ & $0.0974$\\
$0.51$ & $2.2395$ & $0.2341$ & $0.2421$ & $0.1046$ & $0.1081$\\
$0.56$ & $2.2718$ & $0.2746$ & $0.2802$ & $0.1209$ & $0.1234$\\
$0.61$ & $2.3065$ & $0.4029$ & $0.4357$ & $0.1747$ & $0.1889$\\
$0.66$ & $2.3432$ & $0.4852$ & $0.5910$ & $0.2071$ & $0.2522$\\
$0.72$ & $2.3828$ & $0.5850$ & $0.6403$ & $0.2455$ & $0.2687$\\
$0.78$ & $2.4257$ & $0.6840$ & $0.7022$ & $0.2820$ & $0.2895$\\
$0.85$ & $2.4720$ & $0.7115$ & $0.7164$ & $0.2878$ & $0.2898$\\
$0.92$ & $2.5223$ & $0.6193$ & $0.6207$ & $0.2455$ & $0.2461$\\
$1.00$ & $2.5766$ & $0.4711$ & $0.4716$ & $0.1828$ & $0.1830$\\
$1.08$ & $2.6355$ & $0.3374$ & $0.3377$ & $0.1280$ & $0.1281$\\
$1.17$ & $2.6994$ & $0.2360$ & $0.2362$ & $0.0874$ & $0.0875$\\
$1.27$ & $2.7687$ & $0.1645$ & $0.1652$ & $0.0594$ & $0.0597$\\
$1.38$ & $2.8438$ & $0.1246$ & $0.1267$ & $0.0438$ & $0.0446$\\
$1.49$ & $2.9254$ & $0.1679$ & $0.1752$ & $0.0574$ & $0.0599$\\
$1.62$ & $3.0141$ & $0.5956$ & $0.6398$ & $0.1976$ & $0.2123$\\
$1.76$ & $3.1104$ & $0.1385$ & $0.1513$ & $0.0445$ & $0.0486$\\
$1.91$ & $3.2152$ & $0.1655$ & $0.1669$ & $0.0515$ & $0.0519$\\
$2.07$ & $3.3293$ & $0.2778$ & $0.2793$ & $0.0834$ & $0.0839$\\
$2.25$ & $3.4535$ & $0.4913$ & $0.4922$ & $0.1423$ & $0.1425$\\
$2.44$ & $3.5888$ & $0.8948$ & $0.8979$ & $0.2493$ & $0.2502$\\
$2.65$ & $3.7362$ & $1.6572$ & $1.6586$ & $0.4436$ & $0.4439$\\
$2.88$ & $3.8969$ & $3.1903$ & $3.1916$ & $0.8187$ & $0.8190$\\
$3.13$ & $4.0723$ & $6.4196$ & $6.4200$ & $1.5764$ & $1.5765$\\
$3.41$ & $4.2637$ & $13.662$ & $13.662$ & $3.2043$ & $3.2044$\\
$3.70$ & $4.4726$ & $31.038$ & $31.039$ & $6.9397$ & $6.9397$\\
$4.03$ & $4.7008$ & $75.669$ & $75.669$ & $16.097$ & $16.097$\\
$4.39$ & $4.9502$ & $196.14$ & $196.14$ & $39.623$ & $39.623$\\
$4.78$ & $5.2227$ & $521.64$ & $521.64$ & $99.880$ & $99.880$\\
$5.2$ & $5.5207$ & $1259.20$ & $1259.2$ & $228.09$ & $228.09$\\
\hline
\multicolumn{6}{|c|}{DNS} \\
\hline
$z_i$& $b_m(z_i)$ & $\sigma_{b_m}$ run C & $\sigma_{b_m}$ run D & $[\sigma_{b_m}/b_m](z_i)$ run C & $[\sigma_{b_m}/b_m](z_i)$ run D\\
\hline
$0.12$ & $1.9583$ & $0.3322$ & $1.7376$ & $0.1696$ & $0.8873$\\
$0.15$ & $1.9867$ & $0.2946$ & $1.7412$ & $0.1483$ & $0.8764$\\
$0.20$ & $2.0234$ & $0.3076$ & $1.7620$ & $0.1520$ & $0.8708$\\
$0.26$ & $2.0705$ & $0.3320$ & $1.8002$ & $0.1604$ & $0.8694$\\
$0.34$ & $2.1306$ & $0.3735$ & $1.8359$ & $0.1753$ & $0.8617$\\
$0.44$ & $2.2060$ & $0.4630$ & $1.8969$ & $0.2099$ & $0.8599$\\
$0.57$ & $2.3028$ & $0.8437$ & $2.5706$ & $0.3664$ & $1.1163$\\
$0.73$ & $2.4263$ & $2.1998$ & $3.2608$ & $0.9066$ & $1.3439$\\
$0.94$ & $2.5818$ & $7.2202$ & $7.3170$ & $2.7965$ & $2.8341$\\
$1.20$ & $2.7797$ & $15.194$ & $15.327$ & $5.4660$ & $5.5141$\\
$1.53$ & $3.0328$ & $3.8116$ & $6.5099$ & $1.2568$ & $2.1465$\\
$1.96$ & $3.3581$ & $214.23$ & $214.23$ & $63.795$ & $63.796$\\
\end{longtable}
\vspace{-0.2cm}
\begin{figure}[ht!]
\centering
\caption{Bias forecasted errors obtained through run C (red dots for DNS, blue dots for DBH, {\it Planck 2018} \cite{Planck_2018} prior on Cosmology) and run D (red cross for DNS, blue cross for DBH, uniform prior on Cosmology); the ET-like scenario with $k_{nl}^0=0.1\ h\text{Mpc}^{-1}$ is assumed. This plot shows only low $z$, where the difference between the results of the runs is not negligible (see Tab. \ref{tab:res_bias}). Prior on cosmology is relevant only in the DNS case.}\vspace{0.2cm}
\includegraphics[width = 0.9\textwidth]{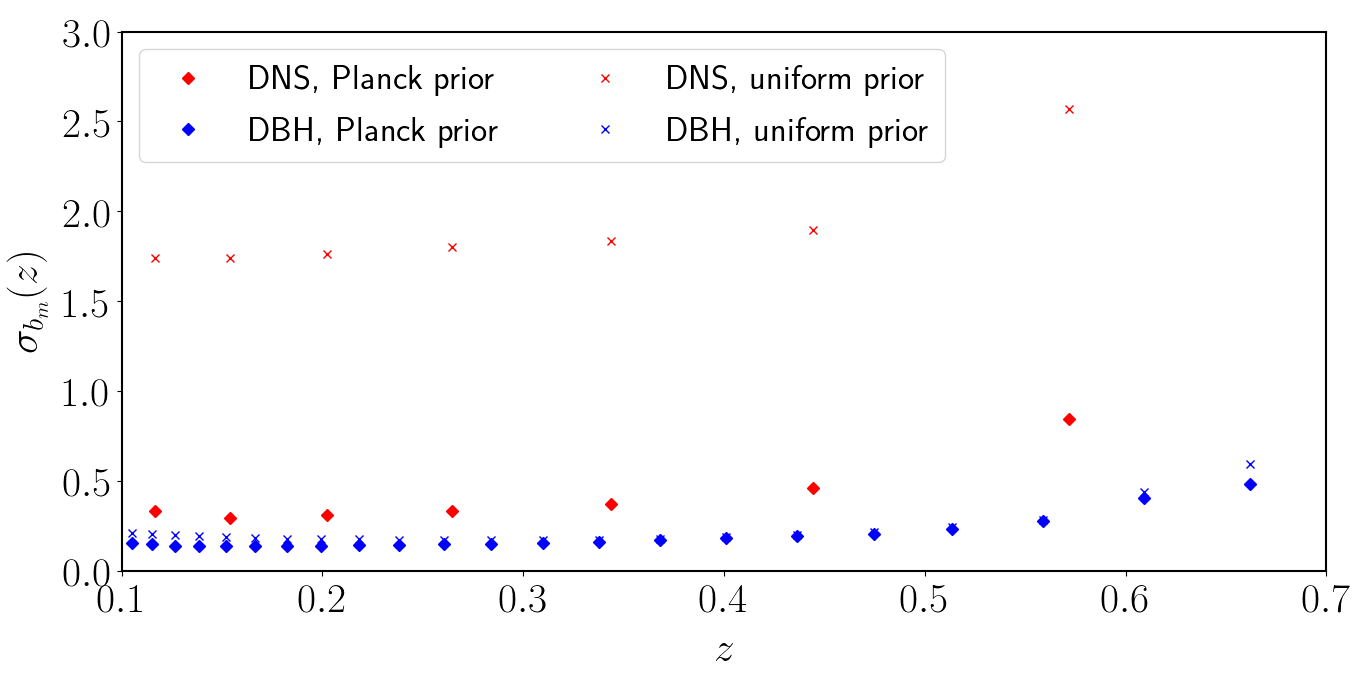}
\vspace{-0.3cm}
\label{fig:bias_no_cosmo}
\end{figure}

\LTcapwidth=\textwidth
\begin{longtable}{|c|c|c|c|c|c|}
\caption{{\small Survey with $k_{nl}^0 = 0.1\ h\text{Mpc}^{-1}$ and $\Delta \Omega = 100\ \text{deg}^2$ separately for DBH and DNS. For each bin from Tab. \ref{tab:DLbin}, the central $z_i$ and the fiducial bias (computed here through eq. (\ref{eq:bias_m})) are indicated. $1\sigma$ marginalized errors and relative errors are shown both for run C and D (see Sec. \ref{sec:bias_res}).}}
\label{tab:res_bias_lin}\\
\hline
\multicolumn{6}{|c|}{DBH}\\
\hline
$z_i$& $b_m(z_i)$ & $\sigma_{b_m}$ run C & $\sigma_{b_m}$ run D & $[\sigma_{b_m}/b_m](z_i)$ run C & $[\sigma_{b_m}/b_m](z_i)$ run D\\
\hline
\endfirsthead
\multicolumn{6}{c}{\footnotesize{Continued from previous page}} \\
$z_i$& $b_m(z_i)$ & $\sigma_{b_m}$ run C & $\sigma_{b_m}$ run D & $[\sigma_{b_m}/b_m](z_i)$ run C & $[\sigma_{b_m}/b_m](z_i)$ run D\\
\hline
\endhead
\multicolumn{6}{c}{\footnotesize{Continued on next page}} \\
\endfoot
\hline
\endlastfoot
$0.10$ & $1.9534$ & $0.2909$ & $0.4416$ & $0.1489$ & $0.2260$\\
$0.12$ & $1.9606$ & $0.2865$ & $0.4371$ & $0.1461$ & $0.2230$\\
$0.13$ & $1.9685$ & $0.2784$ & $0.4325$ & $0.1414$ & $0.2197$\\
$0.14$ & $1.9770$ & $0.2761$ & $0.4286$ & $0.1397$ & $0.2168$\\
$0.15$ & $1.9863$ & $0.2787$ & $0.4241$ & $0.1403$ & $0.2135$\\
$0.17$ & $1.9965$ & $0.2850$ & $0.4189$ & $0.1428$ & $0.2098$\\
$0.18$ & $2.0076$ & $0.2943$ & $0.4132$ & $0.1466$ & $0.2058$\\
$0.20$ & $2.0196$ & $0.3061$ & $0.4072$ & $0.1516$ & $0.2016$\\
$0.22$ & $2.0326$ & $0.3202$ & $0.4020$ & $0.1575$ & $0.1978$\\
$0.24$ & $2.0468$ & $0.3371$ & $0.3990$ & $0.1647$ & $0.1949$\\
$0.26$ & $2.0622$ & $0.3573$ & $0.4008$ & $0.1733$ & $0.1944$\\
$0.28$ & $2.0789$ & $0.3816$ & $0.4102$ & $0.1836$ & $0.1973$\\
$0.31$ & $2.0969$ & $0.4115$ & $0.4301$ & $0.1963$ & $0.2051$\\
$0.34$ & $2.1165$ & $0.4496$ & $0.4636$ & $0.2124$ & $0.2190$\\
$0.37$ & $2.1377$ & $0.4987$ & $0.5131$ & $0.2333$ & $0.2400$\\
$0.40$ & $2.1606$ & $0.5636$ & $0.5816$ & $0.2609$ & $0.2692$\\
$0.44$ & $2.1854$ & $0.6491$ & $0.6729$ & $0.2970$ & $0.3079$\\
$0.47$ & $2.2113$ & $0.7570$ & $0.7950$ & $0.3423$ & $0.3595$\\
$0.51$ & $2.2395$ & $0.9181$ & $0.9510$ & $0.4100$ & $0.4247$\\
$0.56$ & $2.2718$ & $1.1519$ & $1.1716$ & $0.5070$ & $0.5157$\\
$0.61$ & $2.3065$ & $1.7488$ & $1.8738$ & $0.7582$ & $0.8124$\\
$0.66$ & $2.3432$ & $2.0532$ & $2.4862$ & $0.8762$ & $1.0610$\\
$0.72$ & $2.3828$ & $2.1273$ & $2.2848$ & $0.8928$ & $0.9589$\\
$0.78$ & $2.4257$ & $1.7837$ & $1.7934$ & $0.7353$ & $0.7393$\\
$0.85$ & $2.4720$ & $1.3113$ & $1.3115$ & $0.5305$ & $0.5305$\\
$0.92$ & $2.5223$ & $0.9435$ & $0.9436$ & $0.3741$ & $0.3741$\\
$1.00$ & $2.5766$ & $0.6886$ & $0.6887$ & $0.2672$ & $0.2673$\\
$1.08$ & $2.6355$ & $0.5115$ & $0.5120$ & $0.1941$ & $0.1943$\\
$1.17$ & $2.6994$ & $0.3884$ & $0.3895$ & $0.1439$ & $0.1443$\\
$1.27$ & $2.7687$ & $0.3050$ & $0.3078$ & $0.1102$ & $0.1112$\\
$1.38$ & $2.8438$ & $0.2691$ & $0.2757$ & $0.0946$ & $0.0969$\\
$1.49$ & $2.9254$ & $0.4277$ & $0.4582$ & $0.1462$ & $0.1566$\\
$1.62$ & $3.0141$ & $1.4770$ & $1.6906$ & $0.4900$ & $0.5609$\\
$1.76$ & $3.1104$ & $0.3315$ & $0.3873$ & $0.1066$ & $0.1245$\\
$1.91$ & $3.2152$ & $0.3483$ & $0.3585$ & $0.1083$ & $0.1115$\\
$2.07$ & $3.3293$ & $0.5416$ & $0.5488$ & $0.1627$ & $0.1648$\\
$2.25$ & $3.4535$ & $0.9206$ & $0.9237$ & $0.2666$ & $0.2675$\\
$2.44$ & $3.5888$ & $1.6438$ & $1.6622$ & $0.4581$ & $0.4632$\\
$2.65$ & $3.7362$ & $3.0340$ & $3.0358$ & $0.8121$ & $0.8126$\\
$2.88$ & $3.8969$ & $5.8245$ & $5.8280$ & $1.4946$ & $1.4955$\\
$3.13$ & $4.0723$ & $11.649$ & $11.650$ & $2.8604$ & $2.8607$\\
$3.41$ & $4.2637$ & $24.412$ & $24.413$ & $5.7256$ & $5.7258$\\
$3.70$ & $4.4726$ & $53.682$ & $53.682$ & $12.002$ & $12.002$\\
$4.03$ & $4.7008$ & $123.09$ & $123.09$ & $26.185$ & $26.185$\\
$4.39$ & $4.9502$ & $287.86$ & $287.86$ & $58.152$ & $58.152$\\
$4.78$ & $5.2227$ & $666.84$ & $666.84$ & $127.68$ & $127.68$\\
$5.20$ & $5.5207$ & $1530.8$ & $1530.8$ & $277.28$ & $277.28$\\
\hline
\multicolumn{6}{|c|}{DNS}\\
\hline
$z_i$& $b_m(z_i)$ & $\sigma_{b_m}$ run C & $\sigma_{b_m}$ run D & $[\sigma_{b_m}/b_m](z_i)$ run C & $[\sigma_{b_m}/b_m](z_i)$ run D \\
\hline
$0.12$ & $1.9583$ & $1.0725$ & $2.2583$ & $0.5477$ & $1.1532$\\
$0.15$ & $1.9867$ & $1.0975$ & $2.2212$ & $0.5524$ & $1.1180$\\
$0.20$ & $2.0234$ & $1.1384$ & $2.3087$ & $0.5626$ & $1.1410$\\
$0.26$ & $2.0705$ & $1.1550$ & $2.7431$ & $0.5578$ & $1.3249$\\
$0.34$ & $2.1306$ & $1.2018$ & $3.6509$ & $0.5640$ & $1.7135$\\
$0.44$ & $2.2060$ & $1.5401$ & $4.7425$ & $0.6981$ & $2.1498$\\
$0.57$ & $2.3028$ & $3.1893$ & $6.7897$ & $1.3850$ & $2.9485$\\
$0.73$ & $2.4263$ & $8.2299$ & $10.328$ & $3.3919$ & $4.2567$\\
$0.94$ & $2.5818$ & $17.427$ & $17.943$ & $6.7500$ & $6.9497$\\
$1.20$ & $2.7797$ & $17.736$ & $17.898$ & $6.3804$ & $6.4389$\\
$1.53$ & $3.0328$ & $4.8000$ & $9.7857$ & $1.5827$ & $3.2267$\\
$1.96$ & $3.3581$ & $226.44$ & $226.49$ & $67.433$ & $67.445$\\
\end{longtable}

\acknowledgments
The authors thank G. Benevento, D. Bertacca, J. Fonseca, A. Raccanelli and P. Zhang for useful discussions. The authors thank M. Spera for the internal review provided for the LIGO/Virgo Collaboration. We thank the anonymous referee for her/his insightful comments, which helped further refine our analysis.
\noindent ML was supported by the project "Combining Cosmic Microwave Background and Large Scale Structure data: an Integrated Approach for Addressing Fundamental Questions in Cosmology", funded by the MIUR Progetti di Ricerca di Rilevante Interesse Nazionale (PRIN) Bando 2017 - grant 2017YJYZAH. ML's work also acknowledges support by the University of Padova under the STARS Grants programme CoGITO, Cosmology beyond Gaussianity, Inference, Theory and Observations. ML, NB and SM acknowledge partial financial support by ASI Grant No. 2016-24-H.0. MM, YB and NG acknowledge financial support from the European Research Council for the ERC Consolidator grant DEMOBLACK, under contract no. 770017. 
MCA and MM acknowledge financial support from the Austrian National Science Foundation through FWF stand-alone grant P31154-N27. 
DK acknowledge financial support from  the South African Radio Astronomy Observatory (SARAO) and the National Research Foundation (Grant No. 75415).

\bibliography{biblio}

\end{document}